\documentclass[twocolumn]{aastex61}
\submitjournal{ApJ}
\hypersetup{linkcolor=red,citecolor=blue,filecolor=cyan,urlcolor=magenta}
\usepackage{hyperref}
\usepackage{amsmath}
\usepackage{amssymb}
\usepackage{graphicx}
\usepackage[utf8]{inputenc}
\DeclareUnicodeCharacter{2212}{-}
\usepackage{longtable}
\usepackage{natbib}
\usepackage{color}

\newcommand{\jonefive}{\object[2MASS J15213995-3538094]{J1521$-$3538}}
\newcommand{\rpro}{\mbox{{\it r}-process}}
\newcommand{\kms}{km\,s$^{-1}$}
\newcommand{\Teff}{\text{T}_{\rm eff}}

\newcommand{\rproc}{$r$-process }

\newcommand{\ncap}{neutron-capture }
\newcommand{\AB}[2]{\mbox{[#1/#2]}}
\newcommand{\feh}{\AB{Fe}{H}}
\newcommand{\xh}{\AB{X}{H}}
\newcommand{\xfe}[1]{\AB{#1}{Fe}}

\newcommand{\rI}{$r$-I }
\newcommand{\rII}{$r$-II }
\newcommand{\rIII}{$r$-III}
\newcommand{\vmic}{$v_\textrm{micr}$}
\newcommand{\longcain}{2MASS~J15213995$-$3538094}
\newcommand{\cain}{J1521$-$3538}

\newcommand{\ji}{J0334$-$5405}

\usepackage[utf8]{inputenc}

\shorttitle{The very metal-poor $r$-III halo star J1521$-$3538 with $\mbox{[Eu/Fe]} = 2.2$
}
\shortauthors{Cain et al.}
\begin{document}

\title{The $R$-Process Alliance:
\cain, a very metal-poor, \\ extremely $r$-process-enhanced star with $\mbox{[Eu/Fe]} = +2.2$, and the class of \rIII\ stars\footnote{This paper includes data gathered with the 6.5\,m Magellan Telescopes located at Las Campanas Observatory, Chile.
Based on observations obtained at the Gemini Observatory (Prop. IDs: GS-2015A-Q-205), which is operated by the Association of Universities for Research in Astronomy (AURA), Inc., under a cooperative agreement with the National Science Foundation (NSF) on behalf of the Gemini partnership: the NSF (United States), the National Research Council (Canada), CONICYT (Chile), Ministerio de Ciencia, Tecnolog\'{i}a e Innovaci\'{o}n Productiva (Argentina), and Minist\'{e}rio da Ci\^{e}ncia, Tecnologia e Inova\c{c}\~{a}o (Brazil).
Based on observations collected at the European Organisation for Astronomical
Research in the Southern Hemisphere under ESO programme(s) 092.D-0308(A).
}}
\correspondingauthor{Anna Frebel}
\email{afrebel@mit.edu}
\author{Madelyn Cain}
\affiliation{Department of Physics \& Kavli Institute for Astrophysics and Space Research, Massachusetts Institute of Technology, Cambridge, MA 02139, USA}
\author{Anna Frebel}
\affiliation{Department of Physics \& Kavli Institute for Astrophysics and Space Research, Massachusetts Institute of Technology, Cambridge, MA 02139, USA}
\affiliation{Joint Institute for Nuclear Astrophysics - Center for the Evolution of the Elements, USA}
\author{Alexander P. Ji}
\affiliation{The Observatories of the Carnegie Institution of Washington, Pasadena, CA 91101, USA}
\affiliation{Hubble Fellow}
\author{Vinicius M. Placco}
\affiliation{Department of Physics, University of Notre Dame, Notre Dame, IN 46556, USA}
\affiliation{Joint Institute for Nuclear Astrophysics - Center for the Evolution of the Elements, USA}
\author{Rana Ezzeddine}
\affiliation{Joint Institute for Nuclear Astrophysics - Center for the Evolution of the Elements, USA}
\affiliation{Department of Physics \& Kavli Institute for Astrophysics and Space Research, Massachusetts Institute of Technology, Cambridge, MA 02139, USA}
\author{Ian U.\ Roederer}
\affiliation{Department of Astronomy, University of Michigan, Ann Arbor, MI 48109, USA}
\affiliation{Joint Institute for Nuclear Astrophysics - Center for the Evolution of the Elements, USA}
\author{Kohei Hattori}
\affiliation{Department of Astronomy, University of Michigan, Ann Arbor, MI 48109, USA}
\affiliation{Department of Physics, Carnegie Mellon University, Pittsburgh, PA 15213, USA}
\author{Timothy C. Beers}
\affiliation{Department of Physics, University of Notre Dame, Notre Dame, IN 46556, USA}
\affiliation{Joint Institute for Nuclear Astrophysics - Center for the Evolution of the Elements,  USA}
\author{Jorge Mel\'endez}
\affiliation{Instituto de Astronomia,  Geof\'{i}sica e Ci\^{e}ncias Atmosf\'{e}ricas,
Universidade de S\~{a}o Paulo, SP 05508-900, Brazil}
\author{Terese T.\ Hansen}
\affiliation{The Observatories of the Carnegie Institution of Washington, 813 Santa Barbara St., Pasadena, CA 91101, USA}
\author{Charli M.\ Sakari}
\affiliation{Department of Physics \& Astronomy, San Francisco State University, San Francisco, CA 94132, USA}
\affiliation{Joint Institute for Nuclear Astrophysics - Center for the Evolution of the Elements, USA}

\begin{abstract}
We report the discovery of \cain, a bright ($V = 12.2$), very metal-poor ([Fe/H] $= -2.8$) strongly $r$-process enhanced field horizontal branch star, based on a high-resolution, high signal-to-noise Magellan/MIKE spectrum.
\cain\ shows the largest $r$-process element over-abundance in any known $r$-process-enhanced star, with \mbox{[Eu/Fe] $ = +2.2$}, and its chemical abundances of 22 neutron-capture elements closely match the scaled solar \rproc pattern.
\cain\ is also one of few known carbon-enhanced metal-poor stars with $r$-process enhancement (CEMP-$r$ stars), as found after correcting the measured C abundance for the star's evolutionary status.
We propose to extend the existing classification of moderately enhanced ($+0.3\leq\mbox{[Eu/Fe]}\leq +1.0$) $r$-I and strongly $r$-process enhanced ($\mbox{[Eu/Fe]}> +1.0$) $r$-II stars to include an \rIII\ class, for $r$-process stars such as \cain, with $\mbox{[Eu/Fe]}> +2.0$ and $\mbox{[Ba/Eu]} < -0.5$, or
$\ge 100$ times the solar ratio of europium to iron.
Using cosmochronometry, we estimate \cain\ to be $12.5\pm5$\,Gyr and $8.9\pm 5$\,Gyr old, using two different sets of initial production ratios. These ages are based on measurements of the Th line at 4019\,\AA\ and other \rpro\ element abundances.
This is broadly consistent with the old age of a low-mass metal-poor field red horizontal branch star. \jonefive\ likely originated in a low-mass dwarf galaxy that was later accreted by the Milky Way, as evidenced by its highly eccentric orbit.
\end{abstract}

\keywords{nucleosynthesis --- Galaxy: halo --- stars: abundances ---  stars: Population II --- stars: individual (\longcain{})}

\section{Introduction\label{section:intro}}

The atmospheres of $\sim13$-billion-year-old stars reflect the chemical composition of interstellar gas at the time of their birth, supplying details about element formation shortly after the Big Bang. Very metal-poor \mbox{([Fe/H] $< -2.0$)} stars are believed to have formed from gas enriched by only one or a few progenitor supernovae or nucleosynthetic events \citep{Frebel15}. A very small fraction (3--5\%) of these ancient stars formed from gas enriched by the rapid neutron-capture \mbox{($r$-)} process (\citealt{Barklem05, hansen18, sakari18},  R. Ezzeddine et al. 2020, in prep.).
The $r$-process is responsible for producing the heaviest elements ($Z>30$) in the Universe, along with the slow \mbox{($s$-)} neutron-capture process \citep{karakas_lattanzio_2014, gull18}, and the intermediate \mbox{($i$-)} neutron-capture process \citep{dardelet14,  hampel16}.
The observed elemental abundance patterns of metal-poor $r$-process-enhanced stars can thus be used to characterize the yields of $r$-process production events and associated sites in the early Universe.

The $r$-process principally occurs when seed nuclei (e.g., iron-group elements) are bombarded rapidly with neutrons, resulting in neutron-rich, unstable isotopes. A distinct chemical abundance pattern of heavy elements up to and including uranium is created from the radioactive decay of these isotopes. 
Evidence from stars displaying the $r$-process pattern, including the Sun and metal-poor stars (e.g., \citealt{Snedenetal:1996, Hill02, Frebel07, casey17, placco17, holmbeck18, sakari18}), strongly suggests that the pattern is universal across cosmic time, at least for elements Ba to Hf. This behavior has been associated with the main component of the $r$-process (the ``main-$r$-process'')~\citep{truran02}.
On the contrary, variations can exist between lighter elements ($38\leq Z<56$) and the corresponding scaled-solar $r$-process pattern (e.g., \citealt{Barklem05, Roederer14d, Ji16b, ji18, cain18}). These variations might be caused by a ``limited $r$-process" which is postulated to only produce neutron-capture elements lighter than barium~\citep{wanajo01, Travaglio04}.
Additionally, the abundances of the actinide elements Th and U occasionally deviate from the scaled-solar $r$-process pattern in the form of stars displaying an ``actinide-boost'' (e.g., \citealt{mashonkina14, holmbeck18}) or an actinide deficiency (e.g., \citealt{ji18}).

Decades of observational and theoretical efforts have been devoted towards understanding the origins of the $r$-process (see \citealt{frebel18} and references therein).
Evidence from the ``kilonova" electromagnetic counterpart of the merger of a neutron star pair  GW170817 \citep{LIGOGW170817a,LIGOGW170817b} strongly supports neutron star mergers as viable main $r$-process sites  (\citealt{coulter17, drout17, kilpatrick17, shappee17}). Simulations also show that collapsar (the collapse of rapidly rotating massive stars) accretion disks may yield sufficient $r$-process elements to explain a significant contribution to the Universe's main $r$-process enrichment~\citep{surman06, siegel_collapsars_2019}. The limited \rproc may occur during core-collapse supernovae, through a high-entropy neutrino wind (e.g., \citealt{Meyer92,Woosley92,Kratz07,Arcones11,Wanajo13}) or jet-like explosions during magnetorotationally driven core-collapse supernovae \citep{Nishimura15}.

The amount of \rproc enhancement in metal-poor stars can provide constraints on the yields and astrophysical sites of the $r$-process(es) if the mass of the birth gas clouds is known. Observational evidence from the $r$-process-rich ultra-faint dwarf (UFD) galaxy Reticulum~II first suggested that the astrophysical site of the main \rproc was a prolific event such as a neutron star merger \citep{Ji16a,Ji16b, Roederer16a}. The gas dilution mass of Reticulum~II and the level of $r$-process-enhancement from several of its metal-poor stars were used to estimate the yield of the rare $r$-process event that enriched the galaxy, consistent with that of a single neutron star merger.

While $r$-process stars in UFDs can provide constraints on the yields of $r$-process events, UFD stars are particularly difficult to observe due to their faint magnitudes. On the contrary, metal-poor $r$-process-enhanced Galactic halo stars are generally brighter and can be easily observed to obtain a very-high-S/N spectrum for detailed chemical abundance analysis. $R$-process enhancement is also not restricted to any particular evolutionary status, making it possible to use different samples to discover them. Nevertheless, most searches have focused on cooler giants whose absorption lines are stronger and thus easier to measure when elemental abundances are low.

About half of metal-poor $r$-process-enhanced halo stars have been estimated to be accreted from now-destroyed $r$-process UFDs such as Reticulum~II \citep{brauer19}. The natal gas cloud masses from which halo stars originally formed, however, remain unknown, so that $r$-process yields cannot be inferred. Information about their birth sites may instead be inferred from their kinematics~\citep{roederer18d} and chemical abundances. Statistically large samples of ancient $r$-process halo stars can thus help trace the origins of $r$-process enhancement in the Galaxy shortly after the Big Bang.

The goal of the $R$-Process Alliance \citep[RPA;][]{hansen18} is to better understand the $r$-process and its astrophysical production site(s) by increasing the number of known  $r$-process-enhanced stars in the Milky Way. Recent results from the RPA include detailed chemical abundances of both strongly enhanced \rII stars \mbox{([Eu/Fe$]> +1.0$)} and moderately enhanced \mbox{\rI} stars \mbox{($+0.3 \leq$ [Eu/Fe] $\leq +1.0$)} in the Galactic halo (e.g., \citealt{cain18, hansen18, holmbeck18, Roederer2018129, sakari18, sakari19}).
We here present a newly discovered very metal-poor ([Fe/H]$ = -2.80$) red horizontal branch star, \longcain{ }(hereafter \cain), found as part of the ongoing discovery work of the RPA.~
\cain\ is extremely $r$-process-enhanced with [Eu/Fe]$ = +2.2$, the highest value of any  $r$-process-enhanced star known to date. \cain\ is also one of a small but growing number of $r$-process-enhanced horizontal branch stars.
In this paper we describe the chemical abundance pattern of \cain, and introduce a new regime in the parameter space of extreme $r$-process-enhancement in very metal-poor stars.

\section{Observations and Line Measurements\label{section:observation_and_line_measurements}}
\cain, with $\alpha =$ 15:21:39.8, $\delta = -$35:38:08.3, $V = 12.2$, was first identified as a potential metal-poor star from photometry
\citep[based on the criteria of][]{melendez16} and spectroscopy in RAVE
DR5 \citep{kunder17}. The star was followed-up with medium-resolution ($R\sim2,000$) spectroscopy using the ESO New Technology Telescope (EFOSC-2; semester
2014A) and the Gemini South Telescope (GMOS-S; semester 2015A).
Both observing setups used similar gratings ($\sim600$~l~mm$^{\rm{-1}}$) and
slits ($\sim1\farcs$0) in blue setup, to cover at least the wavelength range
$\sim3800$-5300\,{\AA} and exposure times to yield signal-to-noise (S/N) ratios of
${\rm S/N}\sim50$ per pixel at $\sim3930$\,\AA.
Further details on the observations, data reduction, and processing can be
found in \citet{Placco18}.
Atmospheric parameters and carbon abundances were determined from the final, combined spectra using the n-SSPP \citep{beers14,beers17}: \mbox{T$_{\rm eff} =  6029$~K}, \mbox{$\log g = 3.02$},  \mbox{$\feh = -2.96$}, \mbox{$\xfe{C} = +1.34$}. Additional medium resolution stellar parameters are listed in Table~\ref{table:parameters}.

\begin{deluxetable}{lcccccccccc}[!ht]
\tabletypesize{\normalsize}
\tablecaption{Stellar Parameters \label{table:parameters}\\}
\tablehead{
\colhead{Source} &
\colhead{$\Teff$} &
\colhead{$\log$ g} & \colhead{\vmic}&  \colhead{\feh} \\
\colhead{} &
\colhead{[K]} &
\colhead{[cgs]} & \colhead{[\kms]}&  \colhead{}}
\startdata
\multicolumn{5}{c}{High-Resolution Stellar Parameters}\\\hline
LTE  & $5850$& $2.10$  &    $2.65$ &  $-2.80$\\
NLTE & $5780$& $2.75$  &    $2.10$  & $-2.54$\\
\hline
\multicolumn{5}{c}{Medium-Resolution Stellar Parameters}\\
\hline
RAVE DR5 & $5586$& $2.84$  &  ...  & $-2.00$\\
RAVE-on\tablenotemark{a} & $6091$& $3.70$  &  ...  & $-1.47$\\
n-SSPP   & $6029$& $3.02$   & ...  & $-2.96$\\
\enddata
\tablecomments{High-resolution LTE parameters are adopted for our chemical abundance analysis.}
\tablenotetext{a}{\citet{casey2017}}
\end{deluxetable}

\begin{figure*}[!ht]
\begin{center}
  \includegraphics[clip=false,width=1.03\textwidth]{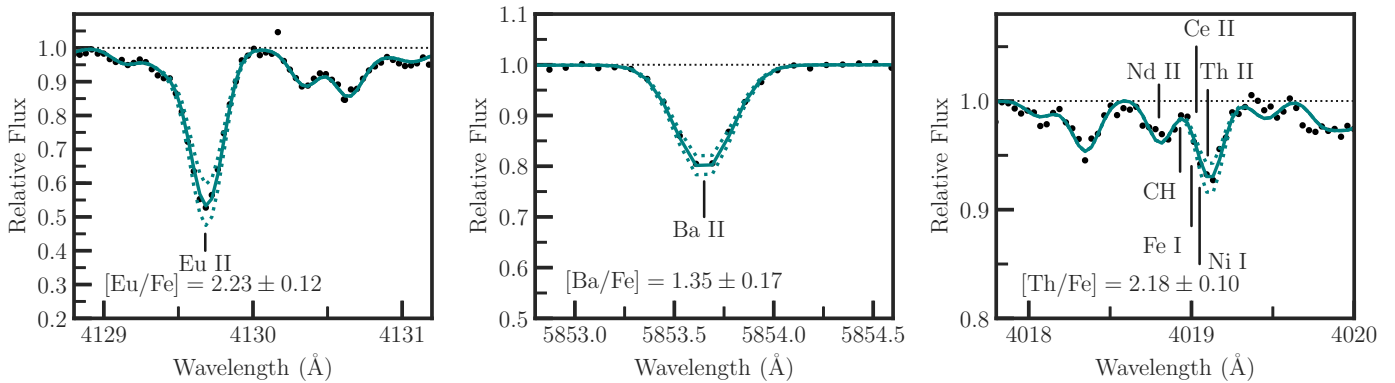}
   \caption{\label{fig:spectrum}
 Portions of the Magellan/MIKE spectrum of \cain\ (black dotted lines) near the Eu\,\textsc{ii} line at 4130\,\AA\ (left), the Ba\,\textsc{ii} line at 5853\,\AA\ (middle), and the Th\,\textsc{ii} line at 4019\,\AA\ (right). Best-fit synthetic spectra are shown in solid teal with abundance variations of $\pm0.14$, $\pm0.10$, and $\pm0.10$\,dex in dashed teal for the Eu\,\textsc{ii}, Ba\,\textsc{ii}, and Th\,\textsc{ii} lines, respectively.}
\end{center}
\end{figure*}

We then observed \cain\  using the Magellan-Clay telescope and the MIKE spectrograph \citep{Bernstein03} at Las Campanas Observatory on 2016 April 16 and 2017 May 6. We obtained a high-resolution spectrum with nominal resolving power of $R \sim35,000$ in the blue and $R \sim28,000$ in the red wavelength regime, using a $0\farcs7$ slit and $2\times2$ binning. The spectra cover $\sim3500$\,{\AA} to $\sim9000$\,{\AA}, with the blue and red CCDs overlapping at around $\sim5000$\,{\AA}. The total exposure time was 15 minutes in 2016 and 20 minutes in 2017.  Data reductions were completed using the MIKE Carnegie Python pipeline \citep{Kelson03}. To combine the data from both nights, we first reduced the data from each night separately. The reduced spectra from both nights were combined after shifting each to the rest frame to account for spatial and spectral shifts between nights. The resulting S/N for the combined spectrum is $110$ at $4000$\,\AA, $150$ at $4500$\,\AA, and $200$ at $6000$\,\AA. Representative portions of the spectrum, including the Eu\,\textsc{ii} line at 4130\,\AA, the Ba\,\textsc{ii} line at 5853\,\AA, and the Th\,\textsc{ii} line at 4019\,\AA, are shown in Figure~\ref{fig:spectrum}.

We measured heliocentric radial velocities (v$_{\mathrm{helio}}$) by cross-correlating the two individual spectra against a template spectrum of HD~140283. Although HD~140283 is a subgiant star, it is an ideal candidate for radial velocity cross-correlation because it has a similar effective temperature and metallicity to \cain\ (see Section~\ref{section:stellar_params}).
We find v$_{\mathrm{helio}}$ values of $+79.32\pm 0.9$ \kms\ and $+81.43 \pm 1.3$\,\kms\ for the 2016 and 2017 observations, respectively. We derive  uncertainties from the standard deviation of v$_{\mathrm{helio}}$ measurements found using several different template spectra. The resulting uncertainties agree with the level of stability of MIKE. We adopt a final heliocentric radial velocity of $80.38$\,\kms\ by averaging our results and an uncertainty of $ \pm 0.84$\,\kms\ by inverse variance weighting.

Previous survey observations also report v$_{\mathrm{helio}}$ values for \cain. RAVE DR5 ~\citep{kunder17} reports a heliocentric radial velocity of $+79.26 \pm 1.84$\,\kms \  from 2005 May 28. Gaia DR2 records a radial velocity of $+79.44 \pm 0.89$\,\kms\ collected between 2014 July 25 and 2016 May 23~\citep{gaiab, gaiaa}. These measurements, along with our data, suggest that \cain\ is a single star, in line with the majority of metal-poor $r$-process-enhanced stars~\citep{Hansen15}.

We then performed a standard abundance analysis for our star following the procedure described in \citet{frebel13a}. We used the 2017 version of the \texttt{MOOG} code\footnote{\href{url}{https://github.com/alexji/moog17scat}} \citep{Sneden73}, which accounts for Rayleigh scattering as coherent, isotropic scattering \citep{Sobeck11}. Together with MOOG, we employ an \texttt{ATLAS9} \citep{Castelli04} 1D plane-parallel model atmosphere with $\alpha$-enhancement and no overshooting, and  assuming local thermodynamic equilibrium (LTE).  All line measurements, stellar parameters, and abundance measurements were made using the \texttt{SMHR} software~\citep{Casey14}. 

We derived iron equivalent widths (EW) using a line list compiled from several data sources (\citealt{obrian91, kurucz_lines, mel09}; \citealt{den14, ruf14}). Neutron-capture line lists used data from \citet{Hill02,hill17}. Synthesis line lists based on atomic data from \citet{Sneden09, sneden14, Sneden16} were provided by Chris Sneden and supplemented with data from \citet{kurucz_lines}. The CH synthesis line list was taken from \citet{Masseron14}.

We obtained EW measurements by fitting  a $\chi^2$-minimized Gaussian profile to each absorption line in our list. Lines with strong damping wings were disregarded. In addition, we performed synthesis measurements for heavily blended lines or lines with hyperfine structure by fitting the line of interest
and any surrounding lines within a local wavelength region by using already measured abundances. We obtained a $3\sigma$ upper limit on the abundance of lines too weak to be detected. [X/H] and \xfe{X} values were calculated using solar abundances from \citet{Asplund09}. Wavelengths ($\lambda$), excitation potentials (EP), $\log gf$ values, EWs, and derived abundances ($\log \epsilon$) are listed in Table~\ref{table:data_short}.

\begin{deluxetable}{crrrrr}
\tablecolumns{6}
\tabletypesize{\tiny}
\tablewidth{0pc}
\tablecaption{\label{table:data_short}
Line list and derived abundances for \cain
}
\tablehead{ \colhead{Element} &
\colhead{$\lambda$} & \colhead{EP} & \colhead{$\log$ \textit{gf}} & \colhead{EW}  & \colhead{$\log\epsilon$(X)}
\\
\colhead{} &
\colhead{[\AA]} & \colhead{[eV]} & \colhead{} & \colhead{[m\AA]}  & \colhead{}
}
\startdata
CH & 4312 & ...  & ...  & syn & 6.20 \\
Na\,\textsc{i} & 5889.95 & 0.00 & 0.11 & 121.0 & 3.69 \\
Na\,\textsc{i} & 5895.92 & 0.00 & $-$0.19 & 89.7 & 3.54 \\
Mg\,\textsc{i} & 3986.75 & 4.35 & $-$1.03 & 12.9 & 5.25 \\
Mg\,\textsc{i} & 4057.51 & 4.35 & $-$0.89 & 11.0 & 5.03 \\
Mg\,\textsc{i} & 4167.27 & 4.35 & $-$0.71 & 23.2 & 5.22 \\
Mg\,\textsc{i} & 4702.99 & 4.33 & $-$0.38 & 35.9 & 5.10 \\
Mg\,\textsc{i} & 5528.40 & 4.34 & $-$0.50 & 32.7 & 5.16 \\
Mg\,\textsc{i} & 5711.09 & 4.34 & $-$1.72 & 3.4 & 5.30 \\
Al\,\textsc{i} & 3944.00 & 0.00 & $-0.64$ & syn & 2.93 \\
Al\,\textsc{i} & 3961.52 & 0.01 & $-$0.34 & 79.4 & 2.88 \\
Si\,\textsc{i} & 3906.52 & 1.91 & $-$1.09 & syn & 4.79 \\
Ca\,\textsc{i}  & 4283.01 & 1.89 & $-$0.22 & 20.2 & 3.86 \\
Ca\,\textsc{i}  & 4318.65 & 1.89 & $-$0.21 & 21.8 & 3.89 \\
\enddata
\tablecomments{``syn" denotes spectrum synthesis was used to measure the abundance. For ease of reading, we include the full list at the end of the paper in Table~\ref{table:data}.}
\end{deluxetable}

\newpage
\section{Stellar parameters\label{section:stellar_params}}
Stellar parameters for \cain\ were determined spectroscopically using EWs from 154 Fe\,\textsc{i} and 21 Fe\,\textsc{ii} lines following the procedure in \citet{frebel13a}. Our final adopted stellar parameters are $\Teff  = 5850$\,K, $\log g = 2.10$, $\feh = -2.80$, and \vmic $ = 2.65$\,\kms.
These parameters are consistent with the star being located on the red horizontal branch, as gleaned from a $12.7$\,Gyr isochrone from the PARSEC models~\citep{Marigo17}\footnote{\href{url}{http://stev.oapd.inaf.it/cgi-bin/cmd}} for $\feh=-2.2$.  
We note here that technically, horizontal branch stars were not included in the \citet{frebel13a} stellar parameter calibration. However, a sample of five horizontal branch stars (A. Frebel et al. 2020, in prep.) suggests that in order to adequately deal with abundances of strong lines, significant effective temperate decreases (together with higher \vmic\ values) are necessary to bring these abundances in line with those of weaker lines. The resulting cooler temperatures are not unlike those investigated in \citet{frebel13a}. While this issue is being explored, we preemptively decided to apply the non-horizontal branch correction to J1512-3538, as a mitigation. 
We adopt typical systematic uncertainties in the stellar parameters as obtained from a spectroscopic analysis \citep{frebel13a,Ji16a}.
We take $\sigma_{\text{T}_{\rm eff}} = 150$\,K, $\sigma_{\log g} = 0.30$\,dex, and $\sigma_{v_\textrm{micr}} = 0.30$\,\kms. Statistical contributions to $\sigma_{\Teff}$, $\sigma_{\log g}$, and $\sigma_{v_\textrm{micr}}$ are negligible in comparison, due to the brightness of \cain\ ($V = 12.2$), the high S/N of the spectrum, and the large number of Fe\,\textsc{i} lines measured. The uncertainty in $\feh$ is $0.15$\,dex, which is derived from the standard deviation of Fe\,\textsc{i} line abundances. Our stellar parameters agree well with results from the EFOSC/GMOS medium-resolution spectrum and results from RAVE surveys (see Table~\ref{table:parameters}).

We also determined stellar parameters assuming non-LTE (NLTE). Deviations from LTE are significant for minority species, such as Fe\,\textsc{i}.  These NLTE effects are particularly strong in metal-poor stars, as they have lower atmospheric electron densities and fewer atomic collisions. Thus, stellar parameters derived assuming NLTE are generally more accurate than the LTE parameters. To determine NLTE stellar parameters, we first determined NLTE abundances for Fe\,\textsc{i} and Fe\,\textsc{ii} lines using a comprehensive Fe atom \citep{Ezzeddine2016} with up-to-date atomic data, especially for hydrogen collisions from \citet{Barklem2018}. Starting from the LTE stellar parameters, we changed each parameter iteratively until excitation and ionization equilibrium were attained in NLTE between abundances of Fe\,\textsc{i} and Fe\,\textsc{ii} lines.  Using this procedure, outlined in \citet{Ezzeddine2017}, we derived NLTE stellar parameters of T$_{\text{eff}}= 5780\pm 112$\,K, $\log g=2.75\pm 0.45$\,dex, $\feh=-2.54\pm$0.15, 
and $v_{\text{micr}}=2.10\pm 0.45$\,\kms. The uncertainties listed are those determined in \citet{ezzeddine16a} for metal-poor stars with detectable Fe\,\textsc{ii} lines, which we adopt here.
The derived temperature and surface gravity are consistent with \cain\ being a red horizontal branch star. The derived NLTE metallicity of $-2.54$ agrees well with the predicted NLTE metallicity of $-2.56$ from the \citet{Ezzeddine2017} equation
$$\feh_{\text{NLTE, predicted}} = 0.86 \times \feh_{\text{LTE}} - 0.15.$$
Since the metallicity was increased and Fe\,\textsc{i} is primarily affected by NLTE, the surface gravity  had to also be increased by $0.65$\,dex to reach ionization balance. The NLTE microturbulence was lowered by 0.55\,\kms.
However, in order to readily calculate consistent abundance ratios and to compare our results to literature values, we use the LTE parameters for the chemical abundance analysis and interpretation throughout the paper. Future capabilities able to produce full NLTE abundance patterns would hopefully make use of our NLTE stellar parameters.

\begin{deluxetable}{lrrrrrrrrrrrrrrrrrr}[!ht]
\tabletypesize{\scriptsize}
\label{table:abund}
\tablewidth{0pc}
\tablecaption{Chemical Abundances of \cain}
\tablehead{
\colhead{Element} &
\colhead{$\log\epsilon (\mbox{X})$} &
\colhead{\xh} &  \colhead{\xfe{X}}& \colhead{N} & \colhead{$\sigma$} & \colhead{}}
\startdata
C\,&$6.20$&$-2.23$&$0.56$&$1$&$0.30$\\
Na\,\textsc{i}&$3.62$&$-2.62$&$0.17$&$2$&$0.10$\\
Mg\,\textsc{i}&$5.14$&$-2.46$&$0.37$&$10$&$0.08$\\
Al\,\textsc{i}&$2.90$&$-3.55$&$-0.76$&$2$&$0.10$\\
Si\,\textsc{i}&$4.79$&$-2.72$&$0.11$&$1$&$0.20$\\
Ca\,\textsc{i}&$3.87$&$-2.47$&$0.32$&$1$&$0.13$\\
Sc\,\textsc{ii}&$0.43$&$-2.72$&$0.07$&$8$&$0.10$\\
Ti\,\textsc{i}&$2.60$&$-2.35$&$0.44$&$8$&$0.16$\\
Ti\,\textsc{ii}&$2.46$&$-2.49$&$0.31$&$37$&$0.13$\\
V\,\textsc{ii}&$1.32$&$-2.12$&$0.18$&$2$&$0.10$\\
Cr\,\textsc{i}&$2.62$&$-3.02$&$-0.23$&$7$&$0.10$\\
Mn\,\textsc{i}&$1.89$&$-3.54$&$-0.75$&$4$&$0.10$\\
Fe\,\textsc{i}&$4.71$&$-2.79$&$0.00$&$154$&$0.15$\\
Fe\,\textsc{ii}&$4.72$&$-2.78$&$0.01$&$21$&$0.10$\\
Co\,\textsc{i}&$2.28$&$-2.71$&$0.08$&$3$&$0.10$\\
Ni\,\textsc{i}&$3.40$&$-2.82$&$-0.03$&$14$&$0.16$\\
Zn\,\textsc{i}&$1.80$&$-2.76$&$0.03$&$1$&$0.20$\\
Sr\,\textsc{ii}&$1.24$&$-1.63$&$1.16$&$1$&$0.20$\\
Y\,\textsc{ii}&$0.45$&$-1.76$&$1.03$&$11$&$0.11$\\
Zr\,\textsc{ii}&$1.13$&$-1.54$&$1.35$&$8$&$0.13$\\
Ru\,\textsc{i}&$0.89$&$-0.86$&$1.93$&$1$&$0.20$\\
Ba\,\textsc{ii}&$0.74$&$-1.44$&$1.35$&$2$&$0.17$\\
La\,\textsc{ii}&$0.12$&$-0.98$&$1.81$&$18$&$0.10$\\
Ce\,\textsc{ii}&$0.43$&$-1.15$&$1.64$&$20$&$0.10$\\
Pr\,\textsc{ii}&$-0.15$&$-0.87$&$1.92$&$8$&$0.10$\\
Nd\,\textsc{ii}&$0.45$&$-0.97$&$1.82$&$46$&$0.10$\\
Sm\,\textsc{ii}&$0.10$&$-0.86$&$1.93$&$13$&$0.10$\\
Eu\,\textsc{ii}&$-0.04$&$-0.56$&$2.23$&$7$&$0.12$\\
Gd\,\textsc{ii}&$0.45$&$-0.62$&$2.17$&$9$&$0.13$\\
Tb\,\textsc{ii}&$-0.37$&$-0.67$&$2.13$&$3$&$0.16$\\
Dy\,\textsc{ii}&$0.45$&$-0.65$&$2.14$&$5$&$0.10$\\
Ho\,\textsc{ii}&$-0.30$&$-0.78$&$2.01$&$9$&$0.10$\\
Er\,\textsc{ii}&$0.21$&$-0.71$&$2.08$&$5$&$0.11$\\
Tm\,\textsc{ii}&$-0.62$&$-0.72$&$2.07$&$5$&$0.15$\\
Yb\,\textsc{ii}&$0.02$&$-0.82$&$1.97$&$1$&$0.20$\\
Hf\,\textsc{ii}&$-0.03$&$-0.88$&$1.91$&$1$&$0.20$\\
Os\,\textsc{i}&$0.90$&$-0.50$&$2.33$&$1$&$0.30$\\
Ir\,\textsc{i}&$<1.53$&$<0.15$&$<2.94$&1&... \\
Th\,\textsc{ii}&$-0.60$&$-0.61$&$2.18$&$1$&$0.10$\\
U\,\textsc{ii}& $<-0.59$ & $<-0.05$ & $<2.75$ & 1 & ... \\
\enddata
\end{deluxetable}

\section{Chemical Abundances}\label{section:abundances}

We obtained abundance measurements using both spectrum synthesis and EW analysis. Our final abundances are summarized in Table~\ref{table:abund} and fully detailed in Table~\ref{table:data_short}. We derive statistical abundance uncertainties, $\sigma$, from the standard deviation of line abundances, which was corrected for small samples when five or fewer lines were measured.
Systematic uncertainties due to, e.g., NLTE effects, 1D stellar model atmospheres, and \textit{gf}-values, are not explicitly considered. Small sample standard deviations for elements with measured abundances of 2--5 lines were obtained following \citet{Keeping62}, by multiplying the range of values covered by our line abundances with the k-factor calculated for small samples.
For elements with one line only, we adopt an uncertainty between 0.1 and 0.3\, dex, depending on the data and fit quality. Finally, we adopt a minimum uncertainty of 0.10\,dex, as better precision is improbable due to continuum placement difficulties.
Table~\ref{table:sys_errors} enumerates the systematic uncertainties in our chemical abundances resulting from uncertainties in our model atmosphere parameters, obtained from varying the stellar parameters by their uncertainties in the positive direction (\mbox{$\sigma_{\Teff} = +150$\,K}, \mbox{$\sigma_{\log g} = +0.30$\,dex}, \mbox{$\sigma_{v_{\text{micr}}} = +0.30$\,\kms}), and recording the resulting change in abundance.

\begin{deluxetable}{lccccccccccccccc}[!ht]
\tabletypesize{\tiny}
\tablewidth{0pc}
\tablecaption{Systematic Uncertainties}
\tablehead{
\colhead{Element} &
\colhead{$\Delta T_\textrm{eff}$} &
\colhead{$\Delta\log(\mbox{g})$} &  \colhead{$\Delta\textrm{v}_\textrm{micr}$}& \colhead{Total} \\
\colhead{} &
\colhead{+150\,K} &
\colhead{+0.30\,dex} &  \colhead{+0.30\,\kms}& \colhead{Error} 
}
\startdata
C\,&+0.10&$-0.28$&$-0.25$&0.38\\
Na\,\textsc{i}&+0.10&$-0.02$&$-$0.07&0.12\\
Mg\,\textsc{i}&+0.07&$-0.05$&$-$0.10&0.13\\
Al\,\textsc{i}&+0.15&$-0.03$&$-$0.01&0.15\\
Ca\,\textsc{i}&+0.09&$-$0.01&$-$0.01&0.11\\
Ti\,\textsc{i}&+0.14&$+$0.00&$+$0.00&0.14\\
Ti\,\textsc{ii}&+0.08&+0.10&$-$0.03&0.13\\
Cr\,\textsc{i}&+0.16&$-$0.01&$-$0.03&0.16\\
Fe\,\textsc{i}&+0.13&$-$0.01&$-$0.04&0.14\\
Fe\,\textsc{ii}&+0.03&+0.10&$-$0.02&0.11\\
Ni\,\textsc{i}&+0.21&$-$0.04&$-$0.08&0.23\\
Zn\,\textsc{i}&+0.10&+0.01&$-$0.01&0.08\\
Sr\,\textsc{ii}&+0.06&+0.13&$+$0.04&0.15\\
Ba\,\textsc{ii}&+0.08&+0.10&$-$0.04&0.13\\
Ce\,\textsc{ii}&$+$0.05&+0.05&$-$0.05&0.09\\
Nd\,\textsc{ii}&+0.11&+0.11&$-$0.01&0.16\\
Eu\,\textsc{ii}&+0.06&+0.07&+0.02&0.09\\
Er\,\textsc{ii}&+0.14&+0.08&$-$0.02&0.16\\
Os\,\textsc{i}&$+0.15$&$+0.14$&$-0.19$&0.28\\
Th\,\textsc{ii}&$+0.07$&$+0.07$&$+0.00$&0.10\\
\enddata
\tablecomments{The individual systematic errors are added in quadrature to compute the total error.}
\end{deluxetable} \label{table:sys_errors}

\begin{figure*}[!ht]
\begin{center}
  \includegraphics[clip=false,width=.9\textwidth]{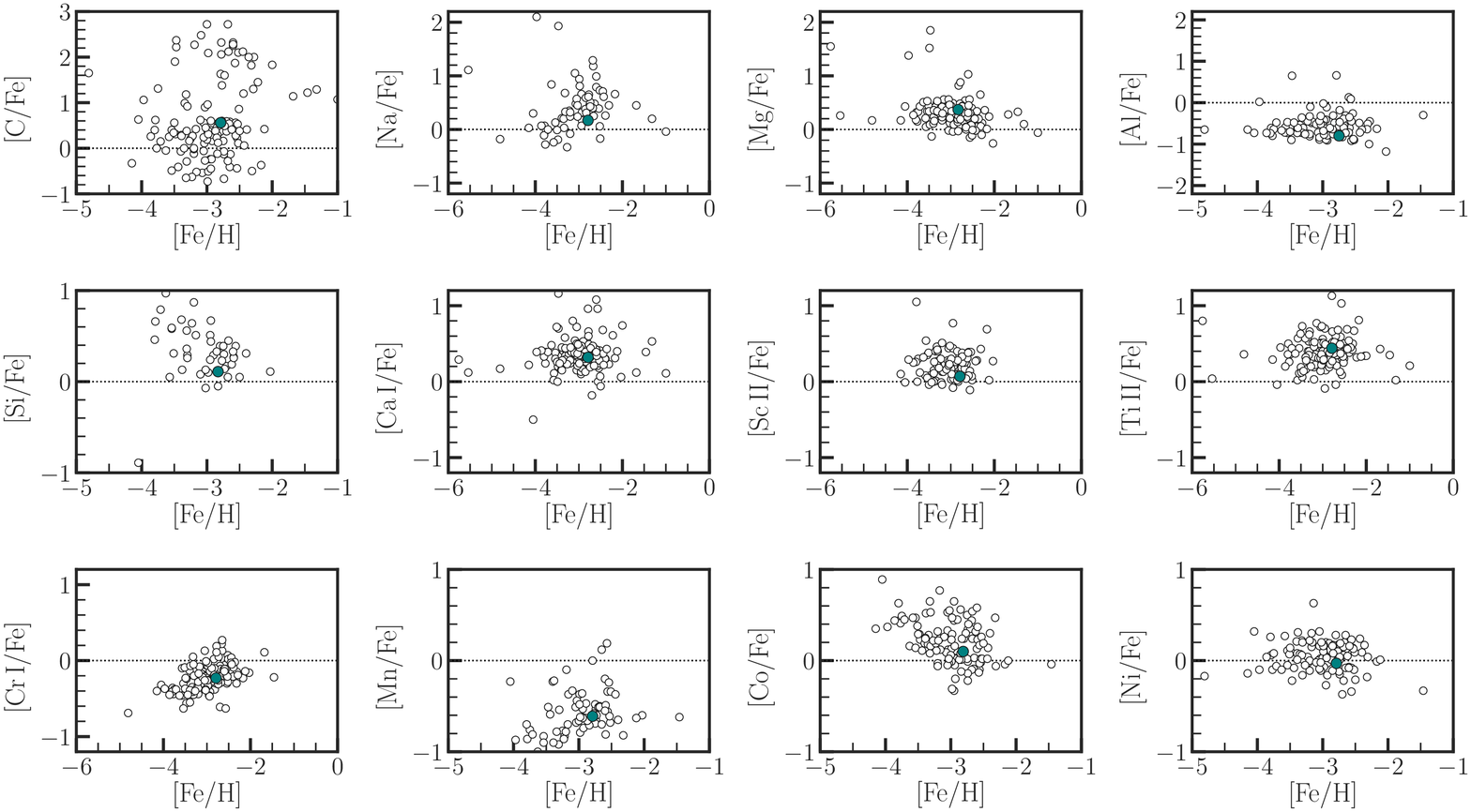}
   \caption{\label{fig:light_elements}
 Light-element abundances for \cain\ (teal filled circles) together with abundances for other metal-poor stars from \citet{yong13_II}. Abundances for C and Na are not corrected for evolutionary status or NLTE behavior in this figure.
 }
\end{center}
\end{figure*}

\subsection{Light Elements}
We measured light-element abundances (C, Na, Mg, Al, Si, Ca, Sc, Ti, V, Cr, Mn, Co, Ni, and Zn) for \cain. They are in agreement with the abundances of other metal-poor stars analyzed by \cite{yong13_II}, as can be seen in Figure~\ref{fig:light_elements}.
We note that we did not measure O (e.g., from the O~\textsc{i} 6300~\AA\ line) or K (e.g., from the K~\textsc{i} 7664~\AA\ line), as they were too weak to be measured, and blended with telluric features.

\textit{Carbon.} We measured a carbon abundance of \mbox{\xfe{C} $= +0.56$} by performing spectrum synthesis on the CH G-bandhead at 4313\,\AA\ and assuming \mbox{\xfe{O} $= +0.00$}. The CH feature at 4323\,\AA\ was too weak to be measured.
We estimate a low value consistent with 4 (the approximate equilibrium value of the CN cycle) for the $^{12}$C/$^{13}$C ratio from the line at 4217\,{\AA}, supporting our assertion that \cain\ is a horizontal branch star.
For the same reason, its carbon abundance can be assumed to be depleted by the CN cycle having operated during the star's evolution along the giant branch.
Therefore, we correct the carbon abundance to account for its evolutionary status to obtain an abundance more representative of the natal gas cloud from which \cain\ formed. We apply the maximum correction of $+0.59$\,dex following \citet{Placco14}, yielding a corrected carbon abundance of \mbox{\xfe{C} $= +1.15$}. \cain\ adds to the small number of known $r$-process rich carbon-enhanced metal-poor stars (CEMP-$r$; with \xfe{C} $> +0.70$; \citealt{Beers05, Aoki07}).

\textit{Sodium and Aluminum.} Sodium abundances (\xfe{Na} $= +0.17$) are derived from the EWs of the Na doublet at 5890\,\AA{ }and 5895\,\AA. There are about seven interstellar medium (ISM) Na features which are clearly spread apart and separate from the stellar lines. It can never be excluded that there is an ISM feature at the same velocity as the star, but since the two Na lines give consistent abundances, it is likely that the stellar lines are not affected. We apply non-LTE corrections to the Na abundance using results from \citet{Lind11}, yielding a corrected value of \mbox{\xfe{Na} $= -0.06$}. We measure the aluminum abundance \mbox{(\xfe{Al} $= -0.76$)} from spectrum synthesis of the Al~I line at 3944\,\AA{ }and an EW measurement of the Al~I at 3961\,\AA. The abundances derived from both lines exhibit excellent agreement, with $\log \epsilon(\rm{Al})$ values of 2.93 and 2.88, respectively.

\textit{Magnesium, Silicon, Calcium, Titanium.}
The abundances of Mg, Si, Ca, and Ti (also known as $\alpha$-elements) were obtained using a mixture of EW measurements and spectrum synthesis.  We measured the Si line at 3905\,\AA{ }using spectrum synthesis. The Si I line at 4102\,\AA\ was heavily blended and could not be measured. The abundances of the $\alpha$-elements were
\mbox{\xfe{Mg} $ = +0.42$}, \mbox{\xfe{Si} $ = +0.11$}, \mbox{\xfe{Ca} $= +0.32$}, and \mbox{\xfe{Ti} $ = +0.31$}.
This level of $\alpha$-element enhancement ($[\alpha/\mathrm{Fe}] \sim+0.4$) is consistent with stars whose light-element abundance enhancement originates primarily from core-collapse supernovae, rather than Type Ia supernovae. The low Si abundance is likely a model atmosphere effect because \cain\ is a hot horizontal branch star (cf.\ \citealt{Preston06}), so the measured abundance likely does not reflect the true cosmic abundance.

\textit{Scandium through Zinc.}
Sc was measured from eight equivalent widths, yielding an abundance of $\mbox{\xfe{Sc} = 0.09}.$
V was measured using spectrum synthesis of the 3952\,\AA{ }line and the 4005\,\AA{ }line. The abundances derived from these lines agreed well, with \mbox{$\xfe{V}$} values of +0.17 and +0.18, respectively.
Ni and Cr were measured from several lines using EW measurements, whereas Mn and Co were measured using synthesis measurements.
Zinc (\mbox{$\xfe{Zn} = +0.03$}) was measured from the EW of the strongest available line at 4810\,\AA. No other Zn lines were detectable. We adopt a conservative error estimate for Zn of $0.20$\,dex, because this line was in fact very weak.

\subsection{Neutron-Capture Elements}
\begin{figure*}[!htb]
\begin{center}
  \includegraphics[clip=false,width=.9\textwidth]{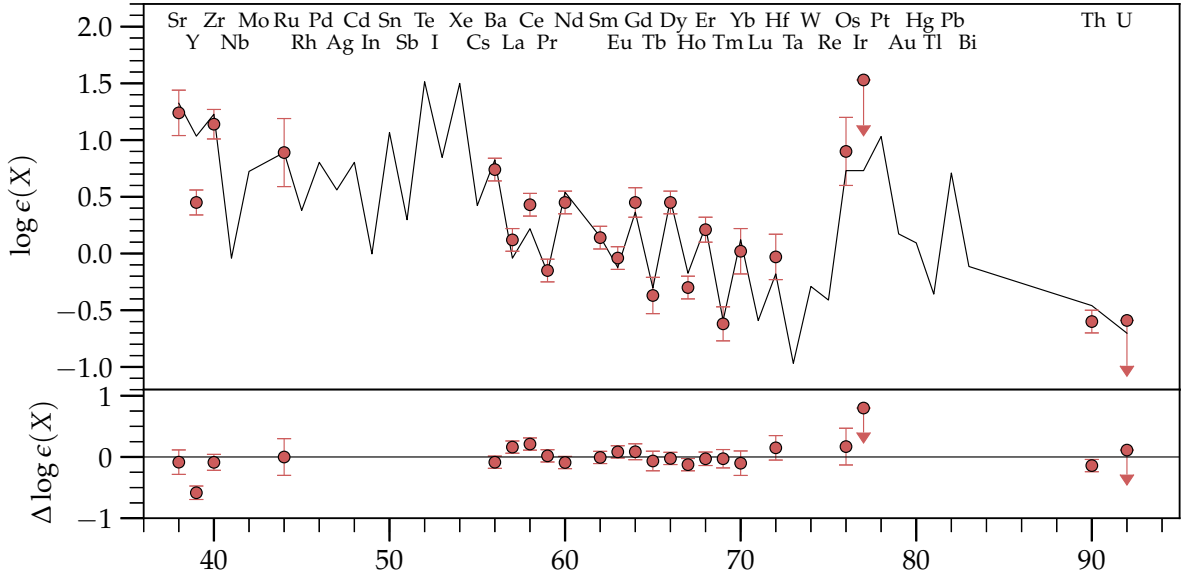}
   \caption{\label{fig:rproc_pattern}
 $R$-process elemental abundance pattern for \cain\ overlaid with a scaled solar $r$-process pattern \citep{Burris00} (top). Residuals between the measured abundances and the scaled solar \rproc pattern are also shown (bottom).
 }
\end{center}
\end{figure*}

We derive abundances for 22 neutron-capture elements for \cain. Measurements or $3\sigma$ upper limits for Sr, Y, Zr, Ru, Ba, La, Pr, Sm, Eu, Tb, Dy, Ho, Tm, Yb, Hf, Os, Ir, Th, and U were measured with spectrum synthesis to account for hyperfine structure and blending of absorption features. Abundances based on EWs were obtained for Ce, Nd, Gd, and Er.  For Ba and Eu measurements, we used the \rproc isotope composition as given in \citet{Sneden08}. Mo, Pd, Ag, Lu, and Pb could not be measured.
Abundance results and uncertainties are given in Table~\ref{table:abund}. The full set of line abundances and associated atomic data of all measured elements are presented in Table~\ref{table:data_short}. Figure~\ref{fig:rproc_pattern} displays our \ncap element abundances overlaid with the scaled solar $r$-process pattern from \citet{Burris00}. We scaled the solar \rproc\ pattern by the difference between the mean measured abundances for elements Ba to Yb in \cain{ }and the Sun.

\textit{Strontium, Yttrium, Zirconium.}
Using spectrum synthesis, we measured the Sr abundance from the line at 4161\,\AA, as the lines at 4077\,\AA{ }and 4215\,\AA{ }were saturated. Y and Zr were measured from 10 and 8 lines, respectively.
These elements are just slightly heavier than those at the first $r$-process peak and may be produced by the limited $r$-process~\citep{wanajo_ishimaru, mello}.
Because the limited \rproc is thought to occur at different astrophysical site(s) than the main $r$-process, the abundances of Sr, Y, and Zr are often offset from the main $r$-process pattern.
In the case of \cain, however, the Sr and Zr abundances agree very well with the scaled solar \rproc pattern, as shown in Figure~\ref{fig:rproc_pattern}. The Y residual is larger ($\sim0.5$\,dex), though
this deviation is common and can be attributed to our choice of solar $r$-process pattern, so it should not be regarded as a concern~\citep{Arlandini99}. The overall agreement between Sr, Y, and Zr and the main \rproc elements with the scaled solar \rproc pattern indicates that \cain{ }likely inherited a substantial amount of material from the main $r$-process.

\textit{Barium through Hafnium.}
Two barium lines were measured using spectrum synthesis at 4130\,\AA{ }and 5853\,\AA. The 4130\,\AA{ }line was blended with Gd and Ce, but a good fit was obtained using the independently measured Gd and Ce abundances.
The line at 5853\,\AA{ }is clean, and is shown in Figure~\ref{fig:spectrum}. Its abundance agreed well with the 4130\,\AA{ }line, yielding an overall uncertainty of $0.11$\,dex. The Ba lines at 4553\,\AA, 6142\,\AA, and 6497\,\AA{ }were saturated. A total of 18 lanthanum lines were measured between 3795\,\AA{ }and 5123\,\AA{ }using spectrum synthesis. All line abundances agreed within 0.3\,dex. Cerium, neodymium, and gadolinium abundances were derived from the EWs of 20, 46, and 9 lines, respectively. The uncertainties on these measurements were low ($\sim0.10$\,dex) due to the large number of lines measured and low spread ($\sim0.30$\,dex).

Eight praseodymium lines were measured between 4063\,\AA{ }and 4449\,\AA. Spectrum synthesis was used to account for the hyperfine structure.  The lines at 4179\,\AA{ }and 4449\,\AA{ }were blended with Nd\,\textsc{ii} and Dy\,\textsc{ii}, respectively. Nevertheless, the derived abundances were consistent with other values, with an overall uncertainty of 0.10\,dex. Samarium was also measured from 13 lines using spectrum synthesis. All of the  measured Sm lines were weak; however, the derived abundance of $\log\epsilon(\rm{Sm}) = 0.10\pm 0.10$
agrees well with the scaled solar \rproc pattern in Figure~\ref{fig:rproc_pattern}.

We measured the europium abundance from seven lines at 3725\,\AA, 3820\,\AA, 3907\,\AA, 4129\,\AA, 4205\,\AA, 6645\,\AA, and 7218\,\AA. The 4129\,\AA\ line is shown in Figure~1.
We used an isotope ratio of $^{151}\rm{Eu}/^{153}\rm{Eu} = 0.88$, following \citet{Sneden08}.
Lines at 6645\,\AA{ }and 7218\,\AA{ }were weaker, whereas the other five lines were strong. The 3725\,\AA{ }line was located in the wing of a Balmer line,
so the continuum was fit locally. The final abundance for the line at 3725\,\AA{ }(with $\log\epsilon(\rm{Eu}) = -0.11$) agreed well with the average line abundance ($\log\epsilon(\rm{Eu}) = -0.04$). The line at 7218\,\AA{ }yielded the highest abundance ($+0.22$) and was included due to its high quality (no blending or distortions). The 4205\,\AA{ }line had the lowest abundance ($-0.16$), but the overall fit was very good. We find \mbox{$\xfe{Eu} = +2.23\pm 0.12$}, with a standard deviation in $\log\epsilon(\rm{Eu})$ values of $0.12$\,dex.

We measured Tb, Dy, Ho, Yb, and Hf using spectrum synthesis, and Er using EW measurements. We measured Tb from three lines at 3569\,\AA, 3659\,\AA, and 3899\,\AA. The line at 3702\,\AA{ }was located in the wing of a Balmer line and could not be measured. Although the 3569\,\AA{ }and 3659\,\AA{ }lines were located in noisier portions of the spectrum, all line measurements agreed well with the average value of \mbox{$\xfe{Tb} = +2.13$}, within $0.20$\,dex. The $\log\epsilon(\rm{Dy})$ measurements were remarkably consistent, with a range of values of $0.05$\,dex. We measured the 4077\,\AA\ Dy line with spectrum synthesis. This required increasing the abundance of the strong blending Sr line by an artificial $\sim0.40$\,dex to account for the saturation of the contributing line. We measured Yb from one line at 3694\,\AA, and adopt a conservative uncertainty of 0.20\,dex. Hf was measured from one line at 4093\,\AA, yielding an abundance of $\log\epsilon(\rm{Hf}) = -0.03$. The Hf line at 3917\,\AA{ }was excluded because it was heavily blended, though we note its abundance ($0.06$) agrees well with our adopted abundance.

Overall, the chemical abundances of barium through hafnium show remarkable agreement with the scaled solar \rproc pattern shown in Figure~\ref{fig:rproc_pattern}.

\textit{Osmium and Iridium.}
We determined the Os abundance from the line at 4419\,\AA{ }using spectrum synthesis. We adopt an uncertainty of 0.30\,dex to account for the quality of the line, which was very weak and heavily blended with Sm. This measurement yielded an abundance of \mbox{$\xfe{Os} = 2.33$} ($\log\epsilon (\rm{Os}) = 0.90$). Our adopted abundance agrees within $1\sigma$ with the scaled solar \rproc pattern.
We also obtained consistent $3\sigma$ upper limits of \mbox{$\log\epsilon (\rm{Os}) = 1.20$} and \mbox{$\log\epsilon (\rm{Os}) = 1.61$} from two weak lines at 4260\,\AA{ }and 4136\,\AA, respectively.

A $3\sigma$ upper limit for Ir was measured from the line at 3513\,\AA, yielding an abundance of $\log\epsilon(\rm{Ir}) < 1.53$. A tighter bound on the abundance upper limit of $\log\epsilon(\rm{Ir}) < 1.13$ was derived from the line at 3800\,\AA, but this was value was not adopted because the line was blended with a Balmer line. Instead, we adopt the more conservative upper limit from the 3513\,\AA{ }line.

\textit{Thorium and Uranium.}
We determined the Th abundance from the line at $4019$\,\AA, taking into account various blends (e.g., C, Fe, Ni, Ce, Pr). The fit, which yielded \mbox{$\xfe{Th} = +2.18$}, is shown in Figure~1. Abundance variations of $\pm 0.10$\,dex are also given to show the quality of the fit.
For U, only an upper limit could be determined from a synthetic fit of the line at 3859\,{\AA}.
This line is located on the wing of a saturated Fe line, so we increased the Fe abundance by $0.49$\,dex to obtain a better fit. We adopt a final conservative, $3\sigma$ upper limit of $\xfe{U} < +2.75$.

Thorium and uranium are radioactive isotopes solely produced by the $r$-process. \cain\ does not appear to have an unusual amount of Th (``actinide boost") compared to expectations of the decay and taking into account its presumably old age. This makes it possible to estimate a stellar age for \cain, by comparing abundance ratios involving Th with theoretical \rproc production ratios, as further described in Section~\ref{section:ages}.

\vspace{3cm}
\section{Discussion and Conclusion\label{section:discussion}}

\subsection{The Class of \rIII\ Stars}

\begin{figure*}[!h]
\begin{center}
  \includegraphics[clip=false,width=.9\textwidth]{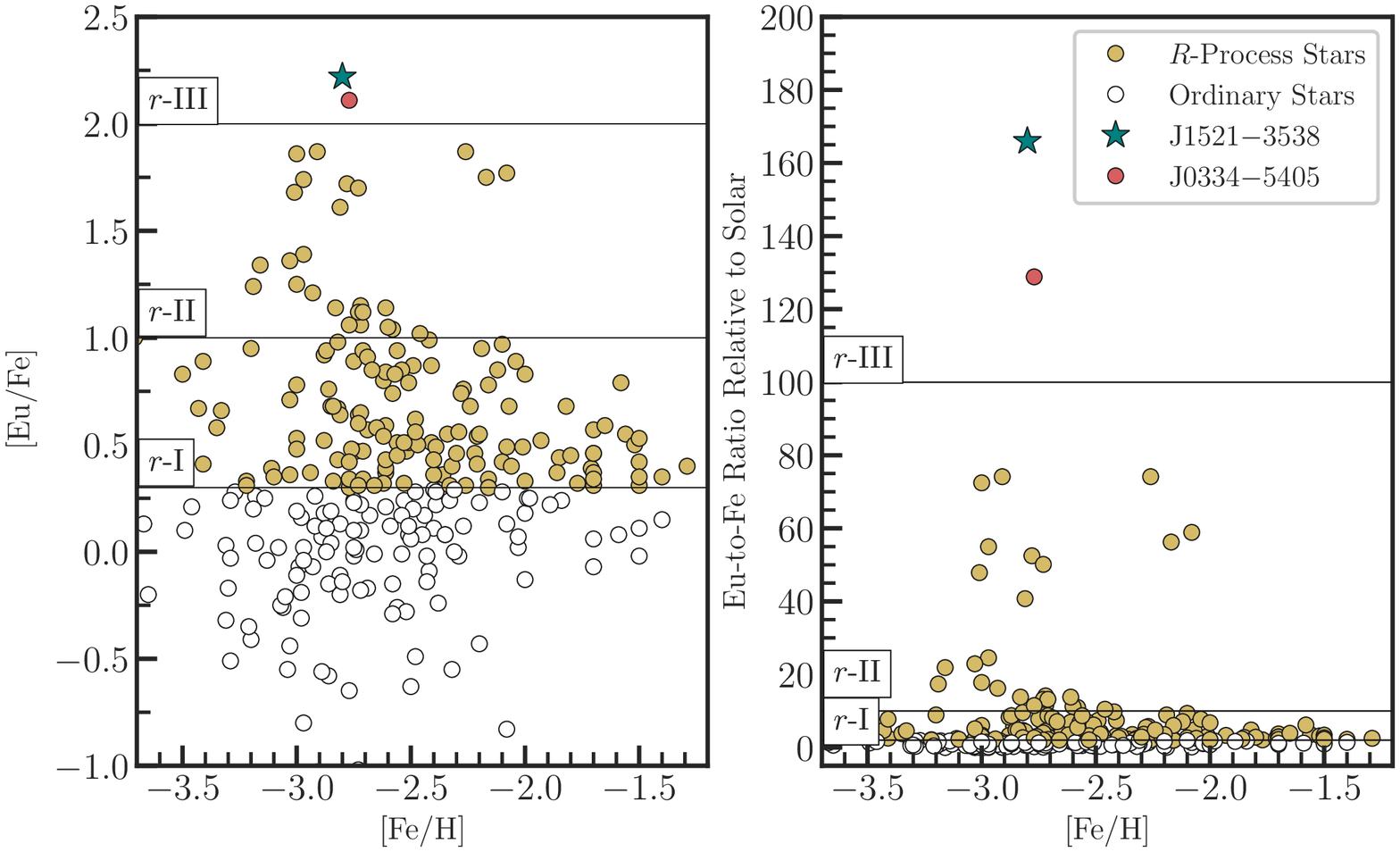}
   \caption{
  \mbox{[Eu/Fe]} and Eu-to-Fe ratio (relative to that of the Sun) plotted against \feh\ for \cain, \ji\ \citep{Ji16a}, and literature star data compiled from the JINAbase for metal-poor stars \citep{Abohalima18}. $R$-I, $r$-II, and \rIII\ boundaries are shown. References are \citet{sku15, roederer10, Burris00, Roederer14c, Roederer14d, hansen15a, Barklem05, johnson2004, Jacobson15, Cayrel04, hayek09, Ji16b, johnson02a, Placco14, roederer08, mello, mello12, cohen06, aoki2002_lp625, aoki02, lai08, spite14, Frebel07, Honda04, honda11, Jonsell06, placco13, ivans_alphapoor, Ivans06, placco15, li15a, McWilliametal, cohen03, barbuy2005, hollek15, masseron2006, preston_sneden01, cui13}.
  See text for discussion. \label{fig:eufe}}
\end{center}
\end{figure*}

\cain\ is the second \rproc enhanced star with a [Eu/Fe] abundance larger than $+2.0$; i.e., the abundance ratio is enhanced by a factor of $>100$ compared to the solar ratio. This is shown in Figure~\ref{fig:eufe}. At $\mbox{[Fe/H]}=-2.8$, this implies that \cain\ contains only 4 times less Eu than the Sun. The other such star known is J0334$-$5405, with $\mbox{[Eu/Fe]} = +2.11$, a star in the $r$-process dwarf galaxy Reticulum\,\textsc{ii} \citep{Ji16a, Roederer16a}.
Though its absolute Eu abundance is very similar to that of \cain, the result was not initially considered significant because the S/N of the spectrum was fairly low.
Given this leap by more than a factor of two in $r$-process enhancement compared to previously known $r$-process rich stars (with $\mbox{[Eu/Fe]} \lesssim+1.9$), we propose to extend the existing classification scheme \citep{Beers05} to include an ``\rIII'' class for stars with [Eu/Fe] $>+2.0$. The other classes are the moderately enhanced ($+0.3\leq\mbox{[Eu/Fe]}\leq +1.0$) \rI and the strongly enhanced $r$-process ($\mbox{[Eu/Fe]}> +1.0$) \rII stars. The three classes are illustrated in Figure~\ref{fig:eufe}. The \rI stars make up about 15\% of metal-poor halo stars with $\mbox{[Fe/H]}\lesssim-2.0$, whereas $r$-II stars are less common with $\sim3$-$5$\% \citep{Barklem05,sakari18}. The \rIII\ stars are thus extremely rare, making up roughly 5\% of all \rII\ stars (excluding J0334$-$5405), which implies a frequency of $<0.3$\% among halo stars.

\subsection{Kinematics \label{section:kinematics}}
The second data release from the Gaia mission
\citep{lindegren18}
included measurements of the
parallax ($\varpi = 0.4449 \pm 0.0614$~mas)
and proper motions
($\mu_{\alpha}\cos\delta = -2.679 \pm 0.088$~mas~yr$^{-1}$,
$\mu_{\delta} = 24.135 \pm 0.062$~mas~yr$^{-1}$)
of \jonefive.
We use these data, plus the coordinates and the heliocentric radial velocity measured from our spectra, to compute the six-dimensional position and space velocity
of \jonefive. We follow the method described by \citet{roederer18d}.
That study adopted \texttt{MWPotential2014}
for the Milky Way gravitational potential \citep{bovy15}, computed orbits using the \texttt{Agama} code \citep{vasiliev19},
and evaluated the specific energy and
integrals of motion using the algorithm described by \citet{binney12},
as implemented in \texttt{Agama}.
We resample the input quantities $10^{3}$ times and recompute the kinematic properties from each resample.
Table~\ref{kinematicstab} lists the results of our calculations.
The uncertainties quoted in Table~\ref{kinematicstab}
reflect statistical uncertainties only,
and they do not account for systematic uncertainties in, e.g.,
the gravitational potential.

Our calculations indicate that \jonefive\ is on a bound,
prograde orbit around the Galaxy, with a high eccentricity of 0.82.
It is currently located relatively near its orbital pericenter
$\sim6$\,kpc from the Galactic center,
and its eccentric orbit carries it 
more than 60\,kpc from the Galactic center and 30\,kpc above the Galactic plane. This presumably makes \jonefive\ an outer halo star.
However,  we note \jonefive\ has energy and actions that are distinct from
any of the 35 highly \rpro-enhanced stars
whose kinematics were studied by \citet{roederer18d}, 
so it is not likely to be affiliated with any of the
groups identified in that study.
All this, taken together with the low metallicity of \jonefive, suggests it would have been
born in a relatively low-mass dwarf galaxy
(e.g., \citealt{kirby13massmetal,walker16}).
The extreme \rpro\ enhancement found in \jonefive\ 
may be a consequence of its birth in such a system if it underwent an early $r$-process event.
A low-mass, low-density dwarf galaxy
on an orbit similar to \jonefive\ would have been
tidally disrupted long ago,
and we suggest that this scenario led to the
accretion of \jonefive\ by the Milky Way.
Searches for other stars with similar kinematic properties
may prove fruitful in the study of
the nature and environment of the \rpro~(e.g., \citealt{yuan20}).

\begin{deluxetable*}{lllc}
\tablecaption{Kinematic properties of J1521$-$3538
\label{kinematicstab}}
\tablewidth{0pt}
\tabletypesize{\small}
\tablehead{
\colhead{Quantity} &
\colhead{Description} &
\colhead{Units} &
\colhead{Value} \\
}
\startdata
$D$             & Distance (inverse parallax)                                & kpc                 & 2.25$^{+0.37}_{-0.29}$      \\
$R_{\rm peri}$  & Orbital pericentric radius                                 & kpc                 & 5.95$^{+0.20}_{-0.21}$      \\
$R_{\rm apo}$   & Orbital apocentric radius                                  & kpc                 & 61.9$^{+53.3}_{-20.2}$      \\
$Z_{\rm max}$   & Maximum height above or below the Galactic midplane        & kpc                 & 32.0$^{+20.9}_{-9.9}$       \\
$e$             & Eccentricity                                               & \nodata             & 0.82$^{+0.08}_{-0.08}$      \\
$V_{R}$         & Radial component of velocity in cylindrical coordinates    & km~s$^{-1}$         & $-$121$^{+10}_{-14}$        \\
$V_{\phi}$      & Azimuthal component of velocity in cylindrical coordinates\tablenotemark{a}
                                                                             & km~s$^{-1}$         & 326$^{+18}_{-16}$           \\
$V_{z}$         & Vertical component of velocity in cylindrical coordinates  & km~s$^{-1}$         & 249$^{+35}_{-27}$           \\
$V_{\perp}$     & $(V_{R}^{2} + V_{z}^{2})^{1/2}$                            & km~s$^{-1}$         & 277$^{+38}_{-29}$           \\
$J_{r}$         & Radial action integral                                     & kpc~km~s$^{-1}$     & 3150$^{+3310}_{-1280}$      \\
$J_{\phi}$      & Azimuthal action integral\tablenotemark{a}
                                                                             & kpc~km~s$^{-1}$     & 2000$^{+20}_{-20}$          \\
$J_{z}$         & Vertical action integral                                   & kpc~km~s$^{-1}$     & 536$^{+107}_{-85}$          \\
$E$             & Specific orbital energy                                    & (km~s$^{-1}$)$^{2}$ & $-$51700$^{+15800}_{-11000}$\\
\enddata
\tablecomments{The uncertainties reflect the range between the
50th percentile and the 84th ($+$) and 16th ($-$) percentiles
of the distributions.}
\tablenotetext{a}{Defined such that this quantity is positive for prograde rotation.}
\end{deluxetable*}

\subsection{Age Estimate for \cain\ \label{section:ages}}

Stellar age estimates can be obtained if radioactive elements are present in the star and their abundances can be measured.
The $r$-process produces thorium and uranium isotopes which have half-lives of 14\,Gyr and 4.7\,Gyr, respectively. Assuming that a single $r$-process event took place soon before the formation of the star, the abundances of such radioactive species can provide information on the age, provided that an initial production ratio is known.
Theoretical \rproc production ratios have been derived \citep{schatz02, hill17} but remain uncertain given our incomplete understanding of the $r$-process itself and its astrophysical site of operation.
Given these production ratios, the age estimate $\Delta t$ can be calculated from
$$\Delta t = 46.78[\log\epsilon(\rm{Th/X}_{\rm{initial}})-\log\epsilon(\rm{Th/X}_{\rm{now}})],$$
originally derived in \citet{Cayreletal:2001}.
Stellar abundance measurement uncertainties also have a significant impact on the resulting age. The statistical uncertainty for ages derived from Th (neglecting uncertainties in the initial production ratio) is given by
$$\sigma_{\Delta t} = 46.78\sqrt{\sigma_{\log\epsilon(\rm{Th})}^2+\sigma_{\log\epsilon(\rm{X})}^2}\label{eq:age_errors},$$
where $\sigma$ is the standard error in abundance.

Despite these challenges, we provide age estimates for \cain\ as derived from elemental abundances of Th with those of other heavy neutron-capture elements, such as Th/Eu. Since U only has an upper limit, it cannot be used to determine a meaningful age or limit. We employ production ratios derived from \rproc\ waiting-point calculations \citep{schatz02} and a high-entropy neutrino wind \rproc model\ \citep{hill17} to illustrate the impact of different theoretical assumptions. We caution that the results should be broadly interpreted only to show that \cain\ and other metal-poor stars are indeed ancient, in line with their low \feh\ values.

Table~\ref{table:ages} list the results for various elemental abundance ratios involving Th for both sets of production ratios. We base our best age estimates on the abundance ratios Th/Ce, Th/Eu, Th/Dy, and Th/Er because they have the lowest statistical uncertainties, following \citet{placco17}.
Our analysis yields 12.5\,Gyr using the \citet{hill17} production ratios and 8.9\,Gyr when using the \citet{schatz02} values. 
We note that an age of 12.5\,Gyr is more in line with expectations for a low-mass metal-poor star with $\feh = -2.8$ on the red horizontal branch. The other age of $\sim9$\,Gyr may also be reasonable, given the uncertainties
and the possibility that \cain{} was born in a low-mass dwarf galaxy.

Age uncertainties stemming from measurement uncertainties are listed in Table~\ref{table:ages}. Realistically, we adopt 5\,Gyr for our age uncertainty.
Uncertainties arising from the production ratios are difficult to quantify but can principally be regarded of order a few Gyr as well.
Historically, just the Th/Eu alone has been used to date many $r$-process stars (e.g., \citealt{johnson01, hayek09}). Using just this ratio, we find 15\,Gyr and 11\,Gyr, respectively.
Regardless of these uncertainties, the results for \cain, as well as the ages of other $r$-process metal-poor stars, show that it remains important to find ways to characterize the progenitor $r$-process event and the birth site of each $r$-process star so that these events can be modeled individually. Given the actinide-boost phenomenon, it has become clear that perhaps not all $r$-process events produce the same amount of $r$-process elements. In this context, understanding the nature and origin of the actinide boost would also be crucial for obtaining more accurate stellar ages via cosmochronometry.

\begin{deluxetable}{lrrrrrll}
\tabletypesize{\scriptsize}
\tablecaption{Stellar age estimates for \cain}
\tablehead{
\colhead{Th$/$X} & \colhead{PR\tablenotemark{i}} & \colhead{Age (Gyr)} & \colhead{PR\tablenotemark{ii}} & \colhead{Age (Gyr)} &\colhead{$\sigma_{\Delta t}$ (Gyr)}}
\startdata
        Th/Ba&$-1.058$&$13.19$&$...$&$... $&$6.29$\\
        Th/La&$-0.362$&$16.74$&$-0.60$&$5.61$&$4.72$\\
Th/Ce&$-0.724$&$14.31$&$-0.79$&$11.22$&$4.72$\\
       Th/Pr&$-0.313$&$6.40$&$-0.30$&$7.02$&$4.78$\\
       Th/Nd&$-0.928$&$5.71$&$-0.91$&$6.55$&$4.72$\\
       Th/Sm&$-0.796$&$-4.49$&$-0.61$&$4.21$&$4.81$\\
Th/Eu&$-0.240$&$14.97$&$-0.33$&$10.76$&$5.21$\\
        Th/Gd&$-0.569$&$22.50$&$-0.81$&$11.23$&$4.97$\\
        Th/Tb&$... $&$... $&$-0.12$&$5.14$&$6.62$\\
Th/Dy&$-0.827$&$10.43$&$-0.89$&$7.48$&$4.71$\\
        Th/Ho&$-0.017$&$13.24$&$... $&$... $&$4.85$\\
Th/Er&$-0.592$&$10.20$&$-0.68$&$6.08$&$5.23$\\
        Th/Tm&$0.155$&$6.31$&$0.12$&$4.68$&$5.71$\\ \hline
Average  & & $12.5$& &  $8.9$  & $\pm 5$ \\
\enddata
\label{table:ages}
\tablenotetext{i}{ Production ratios from  a high-entropy neutrino wind \rproc model \citep{hill17}.}
\tablenotetext{ii}{ Production ratios from \rproc\ waiting-point calculations \citep{schatz02}.}
\tablecomments{Abundance ratios used for determining ages are Th/Ce, Th/Eu, Th/Dy, and Th/Er. Age averages are given $\pm$ the standard deviation of the age measurements included in the average, for illustrative purposes. $\sigma_{\Delta t}$ is the statistical error calculated from measurement uncertainties. The large spread in age estimates reflects measurement uncertainties and uncertainties in the theoretical models used to derive the production ratios.
}
\end{deluxetable}

\section{Summary}
The metal-poor horizontal branch star \cain\ was observed as part of efforts by the $R$-Process Alliance that aim to advance our understanding of the \rproc by studying $r$-process-enhanced Galactic halo stars.
We have discovered that \cain{} displays the strongest over-abundance of \rproc elements observed in any $r$-process-enhanced star.
The heavy-element abundance pattern from Sr to Th closely matches the scaled solar \rproc pattern.
 A kinematic analysis of \jonefive\ shows it to be on a bound, prograde orbit around the Galaxy. This is distinct from other highly $r$-process-enhanced stars \citep{roederer18d}. Given that it also has a highly eccentric orbit, it was likely accreted from a low-mass dwarf galaxy. Other \rproc stars with a similar kinematic signature may confirm this origin scenario.

Our detailed chemical abundance analysis shows that \cain{ } is $\sim12.5$\,Gyr and $\sim8.9$\,Gyr old using the technique of cosmochronometry, which invokes abundance ratios of a radioactive element to other stable \rpro\  elements. 
We used abundances from the Th\,II line at 4019\,\AA\ and lines of various \rpro\ elements to derive these adopted ages.
The two ages result from using two different sets of initial production ratios. The discrepancy highlights that the underlying r-process theory still requires fine tuning, or that the astrophysical site may play a significant role in the elemental yields of actinide-to-stable \rpro\ element abundances. Hopefully, major facilities such as the Facility for Rare Isotope Beams (FRIB) will help to alleviate such discrepancies by providing fundamental nuclear species measurements that ultimately improve \rproc\ predictions.
Regardless, \jonefive\ is clearly an old star, which is supported by it being a low-mass metal-poor field red horizontal branch star.  
Hence, \jonefive\ adds to the growing inventory of highly \rproc enhanced stars that will help to better characterize the nature of the $r$-process and its astrophysical site of operation.
\acknowledgements
We thank Chris Sneden for providing  an up-to-date version of his neutron-capture line lists. M.~C. acknowledges support from the MIT UROP program and the Heising-Simons Foundation.
A.~F. is partially supported by NSF-CAREER grant AST-1255160 and AST-1716251. This work benefited from support by the National Science Foundation under Grant No. PHY-1430152 (JINA Center for the Evolution of the Elements). A.~P.~J. is supported by NASA through Hubble Fellowship grant HST-HF2-51393.001 awarded by the Space Telescope Science Institute, which is operated by the Association of Universities for Research in Astronomy, Inc., for NASA, under contract NAS5-26555. I.~U.~R. acknowledges support from NSF grants AST~1613536 and AST~1815403. K.~H. is supported by NASA-ATP award
NNX15AK79G.
T.~C.~B. acknowledges partial support from the Leverhulme Trust (UK), during his visiting professorship at the University Of Hull, when this paper was finished.
This work made extensive use of NASA's Astrophysics Data System Bibliographic Services.


\facilities{Magellan:Clay (MIKE), Gemini:South (GMOS)}
\software{MOOG~\citep{Sneden73,Sobeck11}, MIKE Carnegie Python Pipeline~\citep{Kelson03}, IRAF~\citep{irafa, irafb}, NumPy~\citep{numpy}, SciPy~\citep{scipy}, Matplotlib~\citep{matplotlib}, Astropy~\citep{astropy}}

\startlongtable
\begin{deluxetable}{crrrrr}
\tablecolumns{6}
\tabletypesize{\tiny}
\tablewidth{0pc}
\tablecaption{\label{table:data}
Line list and derived abundances for \cain
}
\tablehead{ \colhead{Element} &
\colhead{$\lambda$} & \colhead{EP} & \colhead{$\log$ \textit{gf}} & \colhead{EW}  & \colhead{$\log\epsilon$(X)}
\\
\colhead{} &
\colhead{[\AA]} & \colhead{[eV]} & \colhead{} & \colhead{[m\AA]}  & \colhead{}
}
\startdata
CH & 4312 & ...  & ...  & syn & 6.20 \\
Na\,\textsc{i} & 5889.95 & 0.00 & 0.11 & 121.0 & 3.69 \\
Na\,\textsc{i} & 5895.92 & 0.00 & $-$0.19 & 89.7 & 3.54 \\
Mg\,\textsc{i} & 3986.75 & 4.35 & $-$1.03 & 12.9 & 5.25 \\
Mg\,\textsc{i} & 4057.51 & 4.35 & $-$0.89 & 11.0 & 5.03 \\
Mg\,\textsc{i} & 4167.27 & 4.35 & $-$0.71 & 23.2 & 5.22 \\
Mg\,\textsc{i} & 4702.99 & 4.33 & $-$0.38 & 35.9 & 5.10 \\
Mg\,\textsc{i} & 5528.40 & 4.34 & $-$0.50 & 32.7 & 5.16 \\
Mg\,\textsc{i} & 5711.09 & 4.34 & $-$1.72 & 3.4 & 5.30 \\
Al\,\textsc{i} & 3944.00 & 0.00 & $-0.64$ & syn & 2.93 \\
Al\,\textsc{i} & 3961.52 & 0.01 & $-$0.34 & 79.4 & 2.88 \\
Si\,\textsc{i} & 3906.52 & 1.91 & $-$1.09 & syn & 4.79 \\
Ca\,\textsc{i}  & 4283.01 & 1.89 & $-$0.22 & 20.2 & 3.86 \\
Ca\,\textsc{i}  & 4318.65 & 1.89 & $-$0.21 & 21.8 & 3.89 \\
Ca\,\textsc{i}  & 4425.44 & 1.88 & $-$0.36 & 13.8 & 3.79 \\
Ca\,\textsc{i}  & 4434.96 & 1.89 & $-$0.01 & 27.9 & 3.81 \\
Ca\,\textsc{i}  & 4454.78 & 1.90 & 0.26 & 61.1 & 4.07 \\
Ca\,\textsc{i}  & 4455.89 & 1.90 & $-$0.53 & 9.7 & 3.81 \\
Ca\,\textsc{i}  & 5265.56 & 2.52 & $-$0.26 & 11.0 & 4.17 \\
Ca\,\textsc{i}  & 5594.47 & 2.52 & 0.10 & 16.1 & 3.99 \\
Ca\,\textsc{i}  & 5598.49 & 2.52 & $-$0.09 & 6.1 & 3.71 \\
Ca\,\textsc{i}  & 5857.45 & 2.93 & 0.23 & 5.5 & 3.72 \\
Ca\,\textsc{i}  & 6102.72 & 1.88 & $-$0.79 & 7.8 & 3.90 \\
Ca\,\textsc{i}  & 6122.22 & 1.89 & $-$0.32 & 18.5 & 3.85 \\
Ca\,\textsc{i}  & 6162.17 & 1.90 & $-$0.09 & 28.9 & 3.87 \\
Ca\,\textsc{i}  & 6439.07 & 2.52 & 0.47 & 20.1 & 3.71 \\
Sc\,\textsc{ii} & 4246.81 & 0.32 & 0.24 & 107.4 & 0.56 \\
Sc\,\textsc{ii}  & 4314.08 & 0.62 & $-$0.10 & 52.1 & 0.40 \\
Sc\,\textsc{ii}  & 4324.98 & 0.59 & $-$0.44 & 29.2 & 0.36 \\
Sc\,\textsc{ii}  & 4400.38 & 0.60 & $-$0.54 & 26.3 & 0.41 \\
Sc\,\textsc{ii}  & 4415.54 & 0.59 & $-$0.67 & 22.9 & 0.46 \\
Sc\,\textsc{ii}  & 5031.01 & 1.36 & $-$0.40 & 8.0 & 0.39 \\
Sc\,\textsc{ii}  & 5526.77 & 1.77 & 0.02 & 7.6 & 0.32 \\
Sc\,\textsc{ii}  & 5657.89 & 1.51 & $-$0.60 & 5.6 & 0.54 \\
Ti\,\textsc{i}  & 3989.76 & 0.02 & $-$0.13 & 19.8 & 2.68 \\
Ti\,\textsc{i}  & 3998.64 & 0.05 & 0.02 & 13.5 & 2.37 \\
Ti\,\textsc{i}  & 4533.24 & 0.85 & 0.54 & 11.1 & 2.49 \\
Ti\,\textsc{i}  & 4534.78 & 0.83 & 0.35 & 9.8 & 2.61 \\
Ti\,\textsc{i}  & 4981.73 & 0.85 & 0.57 & 9.8 & 2.38 \\
Ti\,\textsc{i}  & 4991.07 & 0.83 & 0.45 & 16.3 & 2.73 \\
Ti\,\textsc{i}  & 4999.50 & 0.82 & 0.32 & 15.3 & 2.82 \\
Ti\,\textsc{i}  & 5007.21 & 0.82 & 0.17 & 9.4 & 2.73 \\
Ti\,\textsc{ii}  & 3489.74 & 0.14 & $-$2.00 & 39.7 & 2.29 \\
Ti\,\textsc{ii}  & 3491.05 & 0.11 & $-$1.10 & 96.0 & 2.40 \\
Ti\,\textsc{ii}  & 3913.46 & 1.11 & $-$0.36 & 124.2 & 2.76 \\
Ti\,\textsc{ii}  & 4025.13 & 0.61 & $-$2.11 & 23.9 & 2.35 \\
Ti\,\textsc{ii}  & 4028.34 & 1.89 & $-$0.92 & 33.0 & 2.59 \\
Ti\,\textsc{ii}  & 4053.82 & 1.89 & $-$1.07 & 10.7 & 2.16 \\
Ti\,\textsc{ii}  & 4163.64 & 2.59 & $-$0.13 & 34.4 & 2.48 \\
Ti\,\textsc{ii}  & 4290.22 & 1.16 & $-$0.87 & 71.3 & 2.35 \\
Ti\,\textsc{ii}  & 4300.04 & 1.18 & $-$0.46 & 102.1 & 2.41 \\
Ti\,\textsc{ii}  & 4337.91 & 1.08 & $-$0.96 & 67.9 & 2.31 \\
Ti\,\textsc{ii}  & 4394.06 & 1.22 & $-$1.77 & 15.9 & 2.37 \\
Ti\,\textsc{ii}  & 4395.03 & 1.08 & $-$0.54 & 112.7 & 2.56 \\
Ti\,\textsc{ii}  & 4395.84 & 1.24 & $-$1.93 & 13.2 & 2.46 \\
Ti\,\textsc{ii}  & 4399.77 & 1.24 & $-$1.19 & 41.7 & 2.33 \\
Ti\,\textsc{ii}  & 4417.71 & 1.17 & $-$1.19 & 56.0 & 2.45 \\
Ti\,\textsc{ii}  & 4418.33 & 1.24 & $-$1.99 & 13.3 & 2.51 \\
Ti\,\textsc{ii}  & 4441.73 & 1.18 & $-$2.41 & 7.1 & 2.58 \\
Ti\,\textsc{ii}  & 4443.80 & 1.08 & $-$0.71 & 97.2 & 2.46 \\
Ti\,\textsc{ii}  & 4450.48 & 1.08 & $-$1.52 & 37.8 & 2.44 \\
Ti\,\textsc{ii}  & 4464.45 & 1.16 & $-$1.81 & 19.1 & 2.43 \\
Ti\,\textsc{ii}  & 4468.49 & 1.13 & $-$0.63 & 103.3 & 2.53 \\
Ti\,\textsc{ii}  & 4501.27 & 1.11 & $-$0.77 & 90.2 & 2.44 \\
Ti\,\textsc{ii}  & 4533.96 & 1.24 & $-$0.53 & 99.2 & 2.46 \\
Ti\,\textsc{ii}  & 4563.77 & 1.22 & $-$0.96 & 78.1 & 2.56 \\
Ti\,\textsc{ii}  & 4571.97 & 1.57 & $-$0.31 & 96.3 & 2.52 \\
Ti\,\textsc{ii}  & 4589.91 & 1.24 & $-$1.79 & 20.3 & 2.51 \\
Ti\,\textsc{ii}  & 4657.20 & 1.24 & $-$2.29 & 8.2 & 2.58 \\
Ti\,\textsc{ii}  & 4708.66 & 1.24 & $-$2.35 & 4.6 & 2.37 \\
Ti\,\textsc{ii}  & 4779.98 & 2.05 & $-$1.37 & 16.5 & 2.77 \\
Ti\,\textsc{ii}  & 4798.53 & 1.08 & $-$2.68 & 5.2 & 2.59 \\
Ti\,\textsc{ii}  & 4805.09 & 2.06 & $-$1.10 & 20.4 & 2.62 \\
Ti\,\textsc{ii}  & 5129.16 & 1.89 & $-$1.34 & 12.4 & 2.43 \\
Ti\,\textsc{ii}  & 5185.90 & 1.89 & $-$1.41 & 8.8 & 2.34 \\
Ti\,\textsc{ii}  & 5188.69 & 1.58 & $-$1.05 & 31.9 & 2.31 \\
Ti\,\textsc{ii}  & 5226.54 & 1.57 & $-$1.26 & 20.7 & 2.28 \\
Ti\,\textsc{ii}  & 5336.79 & 1.58 & $-$1.60 & 14.8 & 2.46 \\
Ti\,\textsc{ii}  & 5381.02 & 1.56 & $-$1.97 & 13.1 & 2.75 \\
V\,\textsc{ii}  & 3951.41 & 1.48 & $-$0.78 & syn & 1.31 \\
V\,\textsc{ii}  & 4005.71 & 1.82 & $-$1.52 & syn & 1.32 \\
Cr\,\textsc{i}  & 3578.68 & 0.00 & 0.42 & 74.9 & 2.56 \\
Cr\,\textsc{i}  & 4254.33 & 0.00 & $-$0.09 & 65.4 & 2.54 \\
Cr\,\textsc{i}  & 4274.80 & 0.00 & $-$0.22 & 58.4 & 2.57 \\
Cr\,\textsc{i}  & 4289.72 & 0.00 & $-$0.37 & 52.7 & 2.64 \\
Cr\,\textsc{i}  & 5206.04 & 0.94 & 0.02 & 28.7 & 2.73 \\
Cr\,\textsc{i}  & 5208.42 & 0.94 & 0.17 & 35.2 & 2.70 \\
Cr\,\textsc{i}  & 5409.77 & 1.03 & $-$0.67 & 4.6 & 2.61 \\
Mn\,\textsc{i} & 4030.75 & 0.00 & $-$0.50 & syn & 1.92 \\
Mn\,\textsc{i} & 4033.06 & 0.00 & $-$0.65 & syn & 1.86 \\
Mn\,\textsc{i} & 4034.48 & 0.00 & $-$0.84 & syn & 1.81 \\
Mn\,\textsc{i} & 4041.35 & 2.11 & 0.28 & syn & 1.98 \\
Fe\,\textsc{i} & 3417.84 & 2.22 & $-$0.68 & 22.6 & 4.55 \\
Fe\,\textsc{i}  & 3476.70 & 0.12 & $-$1.51 & 104.1 & 4.96 \\
Fe\,\textsc{i}  & 3497.84 & 0.11 & $-$1.55 & 95.8 & 4.78 \\
Fe\,\textsc{i}  & 3521.26 & 0.91 & $-$0.99 & 61.4 & 4.29 \\
Fe\,\textsc{i}  & 3558.52 & 0.99 & $-$0.63 & 107.9 & 5.03 \\
Fe\,\textsc{i}  & 3608.86 & 1.01 & $-$0.09 & 114.1 & 4.67 \\
Fe\,\textsc{i}  & 3610.16 & 2.81 & 0.12 & 38.8 & 4.65 \\
Fe\,\textsc{i}  & 3709.25 & 0.91 & $-$0.62 & 110.1 & 4.50 \\
Fe\,\textsc{i}  & 3727.62 & 0.96 & $-$0.61 & 117.8 & 4.69 \\
Fe\,\textsc{i}  & 3742.62 & 2.94 & $-$0.81 & 13.5 & 4.88 \\
Fe\,\textsc{i}  & 3765.54 & 3.24 & 0.48 & 55.6 & 4.70 \\
Fe\,\textsc{i}  & 3787.88 & 1.01 & $-$0.84 & 100.5 & 4.58 \\
Fe\,\textsc{i}  & 3790.09 & 0.99 & $-$1.74 & 39.4 & 4.51 \\
Fe\,\textsc{i}  & 3795.00 & 0.99 & $-$0.74 & 108.5 & 4.63 \\
Fe\,\textsc{i}  & 3805.34 & 3.30 & 0.31 & 66.3 & 5.07 \\
Fe\,\textsc{i}  & 3812.97 & 0.96 & $-$1.05 & 106.4 & 4.86 \\
Fe\,\textsc{i}  & 3816.34 & 2.20 & $-$1.20 & 9.2 & 4.37 \\
Fe\,\textsc{i}  & 3841.05 & 1.61 & $-$0.04 & 113.9 & 4.64 \\
Fe\,\textsc{i}  & 3846.80 & 3.25 & $-$0.02 & 28.9 & 4.77 \\
Fe\,\textsc{i}  & 3849.97 & 1.01 & $-$0.86 & 108.1 & 4.75 \\
Fe\,\textsc{i}  & 3865.52 & 1.01 & $-$0.95 & 103.5 & 4.74 \\
Fe\,\textsc{i}  & 3867.22 & 3.02 & $-$0.45 & 16.2 & 4.67 \\
Fe\,\textsc{i}  & 3872.50 & 0.99 & $-$0.89 & 102.5 & 4.63 \\
Fe\,\textsc{i}  & 3878.02 & 0.96 & $-$0.90 & 107.7 & 4.71 \\
Fe\,\textsc{i}  & 3887.05 & 0.91 & $-$1.14 & 81.1 & 4.43 \\
Fe\,\textsc{i}  & 3895.66 & 0.11 & $-$1.67 & 112.8 & 4.75 \\
Fe\,\textsc{i}  & 3898.01 & 1.01 & $-$2.04 & 39.6 & 4.82 \\
Fe\,\textsc{i}  & 3902.95 & 1.56 & $-$0.44 & 93.7 & 4.58 \\
Fe\,\textsc{i}  & 3917.18 & 0.99 & $-$2.15 & 23.5 & 4.62 \\
Fe\,\textsc{i}  & 3920.26 & 0.12 & $-$1.73 & 108.5 & 4.73 \\
Fe\,\textsc{i}  & 3940.88 & 0.96 & $-$2.60 & 11.7 & 4.68 \\
Fe\,\textsc{i}  & 3949.95 & 2.18 & $-$1.25 & 10.9 & 4.48 \\
Fe\,\textsc{i}  & 3977.74 & 2.20 & $-$1.12 & 19.2 & 4.64 \\
Fe\,\textsc{i}  & 4005.24 & 1.56 & $-$0.58 & 92.9 & 4.68 \\
Fe\,\textsc{i}  & 4009.71 & 2.22 & $-$1.25 & 14.6 & 4.65 \\
Fe\,\textsc{i}  & 4014.53 & 3.05 & $-$0.59 & 21.2 & 4.97 \\
Fe\,\textsc{i}  & 4067.98 & 3.21 & $-$0.53 & 15.0 & 4.88 \\
Fe\,\textsc{i}  & 4071.74 & 1.61 & $-$0.01 & 120.8 & 4.68 \\
Fe\,\textsc{i}  & 4076.63 & 3.21 & $-$0.59 & 19.8 & 5.08 \\
Fe\,\textsc{i}  & 4084.49 & 3.33 & $-$0.54 & 7.3 & 4.66 \\
Fe\,\textsc{i}  & 4132.06 & 1.61 & $-$0.68 & 96.1 & 4.86 \\
Fe\,\textsc{i}  & 4132.90 & 2.85 & $-$1.01 & 8.6 & 4.75 \\
Fe\,\textsc{i}  & 4134.68 & 2.83 & $-$0.65 & 19.4 & 4.77 \\
Fe\,\textsc{i}  & 4137.00 & 3.42 & $-$0.45 & 10.3 & 4.81 \\
Fe\,\textsc{i}  & 4143.87 & 1.56 & $-$0.51 & 97.5 & 4.67 \\
Fe\,\textsc{i}  & 4147.67 & 1.48 & $-$2.07 & 13.5 & 4.71 \\
Fe\,\textsc{i}  & 4153.90 & 3.40 & $-$0.28 & 9.9 & 4.61 \\
Fe\,\textsc{i}  & 4156.80 & 2.83 & $-$0.81 & 13.1 & 4.73 \\
Fe\,\textsc{i}  & 4157.78 & 3.42 & $-$0.40 & 10.9 & 4.79 \\
Fe\,\textsc{i}  & 4172.75 & 0.96 & $-$3.02 & 5.2 & 4.70 \\
Fe\,\textsc{i}  & 4175.64 & 2.85 & $-$0.83 & 21.2 & 5.00 \\
Fe\,\textsc{i}  & 4181.76 & 2.83 & $-$0.37 & 29.3 & 4.70 \\
Fe\,\textsc{i}  & 4184.89 & 2.83 & $-$0.87 & 12.0 & 4.75 \\
Fe\,\textsc{i}  & 4187.04 & 2.45 & $-$0.56 & 39.2 & 4.70 \\
Fe\,\textsc{i}  & 4187.80 & 2.42 & $-$0.51 & 37.1 & 4.59 \\
Fe\,\textsc{i}  & 4191.43 & 2.47 & $-$0.67 & 33.7 & 4.74 \\
Fe\,\textsc{i}  & 4199.10 & 3.05 & 0.16 & 43.0 & 4.62 \\
Fe\,\textsc{i}  & 4202.03 & 1.49 & $-$0.69 & 92.7 & 4.68 \\
Fe\,\textsc{i}  & 4216.18 & 0.00 & $-$3.36 & 16.8 & 4.64 \\
Fe\,\textsc{i}  & 4217.55 & 3.43 & $-$0.48 & 11.5 & 4.91 \\
Fe\,\textsc{i}  & 4222.21 & 2.45 & $-$0.91 & 17.7 & 4.61 \\
Fe\,\textsc{i}  & 4227.43 & 3.33 & 0.27 & 41.3 & 4.75 \\
Fe\,\textsc{i}  & 4233.60 & 2.48 & $-$0.60 & 30.0 & 4.61 \\
Fe\,\textsc{i}  & 4238.81 & 3.40 & $-$0.23 & 16.1 & 4.79 \\
Fe\,\textsc{i}  & 4250.12 & 2.47 & $-$0.38 & 45.3 & 4.63 \\
Fe\,\textsc{i}  & 4250.79 & 1.56 & $-$0.71 & 86.3 & 4.67 \\
Fe\,\textsc{i}  & 4260.47 & 2.40 & 0.08 & 80.9 & 4.62 \\
Fe\,\textsc{i}  & 4271.15 & 2.45 & $-$0.34 & 48.3 & 4.61 \\
Fe\,\textsc{i}  & 4271.76 & 1.49 & $-$0.17 & 130.9 & 4.89 \\
Fe\,\textsc{i}  & 4282.40 & 2.18 & $-$0.78 & 46.6 & 4.76 \\
Fe\,\textsc{i}  & 4325.76 & 1.61 & 0.01 & 133.0 & 4.87 \\
Fe\,\textsc{i}  & 4352.73 & 2.22 & $-$1.29 & 15.8 & 4.70 \\
Fe\,\textsc{i}  & 4375.93 & 0.00 & $-$3.00 & 37.7 & 4.71 \\
Fe\,\textsc{i}  & 4404.75 & 1.56 & $-$0.15 & 127.4 & 4.84 \\
Fe\,\textsc{i}  & 4415.12 & 1.61 & $-$0.62 & 90.0 & 4.66 \\
Fe\,\textsc{i}  & 4427.31 & 0.05 & $-$2.92 & 37.2 & 4.67 \\
Fe\,\textsc{i}  & 4442.34 & 2.20 & $-$1.23 & 18.4 & 4.69 \\
Fe\,\textsc{i}  & 4447.72 & 2.22 & $-$1.34 & 12.5 & 4.63 \\
Fe\,\textsc{i}  & 4459.12 & 2.18 & $-$1.28 & 15.8 & 4.64 \\
Fe\,\textsc{i}  & 4461.65 & 0.09 & $-$3.19 & 22.0 & 4.68 \\
Fe\,\textsc{i}  & 4466.55 & 2.83 & $-$0.60 & 19.4 & 4.69 \\
Fe\,\textsc{i}  & 4476.02 & 2.85 & $-$0.82 & 13.0 & 4.73 \\
Fe\,\textsc{i}  & 4489.74 & 0.12 & $-$3.90 & 2.4 & 4.39 \\
Fe\,\textsc{i}  & 4494.56 & 2.20 & $-$1.14 & 20.5 & 4.66 \\
Fe\,\textsc{i}  & 4528.61 & 2.18 & $-$0.82 & 44.2 & 4.75 \\
Fe\,\textsc{i}  & 4531.15 & 1.48 & $-$2.10 & 12.6 & 4.67 \\
Fe\,\textsc{i}  & 4602.94 & 1.49 & $-$2.21 & 8.8 & 4.61 \\
Fe\,\textsc{i}  & 4736.77 & 3.21 & $-$0.67 & 9.2 & 4.75 \\
Fe\,\textsc{i}  & 4871.32 & 2.87 & $-$0.34 & 23.0 & 4.54 \\
Fe\,\textsc{i}  & 4872.14 & 2.88 & $-$0.57 & 16.7 & 4.62 \\
Fe\,\textsc{i}  & 4890.76 & 2.88 & $-$0.38 & 30.7 & 4.75 \\
Fe\,\textsc{i}  & 4891.49 & 2.85 & $-$0.11 & 37.2 & 4.57 \\
Fe\,\textsc{i}  & 4903.31 & 2.88 & $-$0.89 & 8.7 & 4.62 \\
Fe\,\textsc{i}  & 4918.99 & 2.85 & $-$0.34 & 23.9 & 4.54 \\
Fe\,\textsc{i}  & 4920.50 & 2.83 & 0.07 & 55.9 & 4.64 \\
Fe\,\textsc{i}  & 4938.81 & 2.88 & $-$1.08 & 4.6 & 4.51 \\
Fe\,\textsc{i}  & 4994.13 & 0.92 & $-$2.97 & 12.1 & 4.94 \\
Fe\,\textsc{i}  & 5001.87 & 3.88 & $-$0.01 & 10.0 & 4.76 \\
Fe\,\textsc{i}  & 5006.12 & 2.83 & $-$0.61 & 16.7 & 4.61 \\
Fe\,\textsc{i}  & 5012.07 & 0.86 & $-$2.64 & 17.4 & 4.73 \\
Fe\,\textsc{i}  & 5041.07 & 0.96 & $-$3.09 & 9.1 & 4.96 \\
Fe\,\textsc{i}  & 5049.82 & 2.28 & $-$1.35 & 13.0 & 4.69 \\
Fe\,\textsc{i}  & 5079.74 & 0.99 & $-$3.25 & 7.4 & 5.05 \\
Fe\,\textsc{i}  & 5083.34 & 0.96 & $-$2.84 & 6.2 & 4.54 \\
Fe\,\textsc{i}  & 5110.41 & 0.00 & $-$3.76 & 13.3 & 4.86 \\
Fe\,\textsc{i}  & 5123.72 & 1.01 & $-$3.06 & 13.9 & 5.18 \\
Fe\,\textsc{i}  & 5133.69 & 4.17 & 0.36 & 9.8 & 4.66 \\
Fe\,\textsc{i}  & 5151.91 & 1.01 & $-$3.32 & 3.6 & 4.82 \\
Fe\,\textsc{i}  & 5162.27 & 4.18 & 0.02 & 5.9 & 4.76 \\
Fe\,\textsc{i}  & 5171.60 & 1.49 & $-$1.72 & 23.0 & 4.56 \\
Fe\,\textsc{i}  & 5191.45 & 3.04 & $-$0.55 & 34.2 & 5.13 \\
Fe\,\textsc{i}  & 5227.19 & 1.56 & $-$1.23 & 53.7 & 4.65 \\
Fe\,\textsc{i}  & 5232.94 & 2.94 & $-$0.06 & 38.5 & 4.61 \\
Fe\,\textsc{i}  & 5266.56 & 3.00 & $-$0.39 & 18.7 & 4.59 \\
Fe\,\textsc{i}  & 5269.54 & 0.86 & $-$1.33 & 96.6 & 4.67 \\
Fe\,\textsc{i}  & 5281.79 & 3.04 & $-$0.83 & 9.1 & 4.73 \\
Fe\,\textsc{i}  & 5302.30 & 3.28 & $-$0.73 & 10.2 & 4.91 \\
Fe\,\textsc{i}  & 5324.18 & 3.21 & $-$0.11 & 22.1 & 4.60 \\
Fe\,\textsc{i}  & 5328.04 & 0.92 & $-$1.47 & 83.2 & 4.65 \\
Fe\,\textsc{i}  & 5328.53 & 1.56 & $-$1.85 & 25.7 & 4.81 \\
Fe\,\textsc{i}  & 5339.93 & 3.27 & $-$0.63 & 5.8 & 4.54 \\
Fe\,\textsc{i}  & 5364.87 & 4.45 & 0.23 & 4.6 & 4.70 \\
Fe\,\textsc{i}  & 5367.47 & 4.42 & 0.44 & 6.2 & 4.59 \\
Fe\,\textsc{i}  & 5369.96 & 4.37 & 0.54 & 11.8 & 4.75 \\
Fe\,\textsc{i}  & 5371.49 & 0.96 & $-$1.64 & 70.4 & 4.69 \\
Fe\,\textsc{i}  & 5383.37 & 4.31 & 0.65 & 16.1 & 4.74 \\
Fe\,\textsc{i}  & 5393.17 & 3.24 & $-$0.91 & 7.7 & 4.92 \\
Fe\,\textsc{i}  & 5397.13 & 0.92 & $-$1.98 & 44.9 & 4.64 \\
Fe\,\textsc{i}  & 5405.77 & 0.99 & $-$1.85 & 48.4 & 4.63 \\
Fe\,\textsc{i}  & 5410.91 & 4.47 & 0.40 & 7.6 & 4.77 \\
Fe\,\textsc{i}  & 5415.20 & 4.39 & 0.64 & 13.1 & 4.71 \\
Fe\,\textsc{i}  & 5424.07 & 4.32 & 0.52 & 14.6 & 4.82 \\
Fe\,\textsc{i}  & 5429.70 & 0.96 & $-$1.88 & 50.9 & 4.67 \\
Fe\,\textsc{i}  & 5434.52 & 1.01 & $-$2.13 & 30.0 & 4.63 \\
Fe\,\textsc{i}  & 5446.92 & 0.99 & $-$1.91 & 46.2 & 4.66 \\
Fe\,\textsc{i}  & 5497.52 & 1.01 & $-$2.83 & 14.5 & 4.95 \\
Fe\,\textsc{i}  & 5506.78 & 0.99 & $-$2.79 & 12.9 & 4.83 \\
Fe\,\textsc{i}  & 5572.84 & 3.40 & $-$0.28 & 12.8 & 4.67 \\
Fe\,\textsc{i}  & 5586.76 & 3.37 & $-$0.11 & 14.8 & 4.54 \\
Fe\,\textsc{i}  & 5615.64 & 3.33 & 0.04 & 21.4 & 4.54 \\
Fe\,\textsc{i}  & 5624.54 & 3.42 & $-$0.76 & 3.4 & 4.56 \\
Fe\,\textsc{i}  & 5709.38 & 3.37 & $-$1.01 & 5.2 & 4.95 \\
Fe\,\textsc{i}  & 6065.48 & 2.61 & $-$1.41 & 5.9 & 4.66 \\
Fe\,\textsc{i}  & 6136.61 & 2.45 & $-$1.41 & 6.8 & 4.57 \\
Fe\,\textsc{i}  & 6191.56 & 2.43 & $-$1.42 & 7.7 & 4.61 \\
Fe\,\textsc{i}  & 6230.72 & 2.56 & $-$1.28 & 6.7 & 4.53 \\
Fe\,\textsc{i}  & 6252.56 & 2.40 & $-$1.69 & 6.3 & 4.76 \\
Fe\,\textsc{i}  & 6393.60 & 2.43 & $-$1.58 & 4.7 & 4.54 \\
Fe\,\textsc{i}  & 6400.00 & 3.60 & $-$0.27 & 8.7 & 4.65 \\
Fe\,\textsc{i}  & 6592.91 & 2.73 & $-$1.47 & 7.9 & 4.95 \\
Fe\,\textsc{i}  & 6677.99 & 2.69 & $-$1.42 & 6.0 & 4.73 \\
Fe\,\textsc{i}  & 7495.07 & 4.22 & $-$0.10 & 4.7 & 4.76 \\
Fe\,\textsc{i}  & 8387.77 & 2.17 & $-$1.51 & 18.6 & 4.76 \\
Fe\,\textsc{i}  & 8688.62 & 2.17 & $-$1.20 & 50.9 & 5.01 \\
Fe\,\textsc{ii}  & 4178.86 & 2.58 & $-$2.51 & 31.7 & 4.72 \\
Fe\,\textsc{ii}  & 4233.17 & 2.58 & $-$1.97 & 70.7 & 4.75 \\
Fe\,\textsc{ii}  & 4416.82 & 2.78 & $-$2.65 & 24.0 & 4.88 \\
Fe\,\textsc{ii}  & 4489.19 & 2.83 & $-$2.96 & 10.2 & 4.80 \\
Fe\,\textsc{ii}  & 4491.41 & 2.86 & $-$2.71 & 13.8 & 4.73 \\
Fe\,\textsc{ii}  & 4508.28 & 2.86 & $-$2.44 & 25.4 & 4.77 \\
Fe\,\textsc{ii}  & 4515.34 & 2.84 & $-$2.60 & 20.9 & 4.81 \\
Fe\,\textsc{ii}  & 4520.22 & 2.81 & $-$2.65 & 18.6 & 4.77 \\
Fe\,\textsc{ii}  & 4555.89 & 2.83 & $-$2.40 & Fe\,\textsc{i}  & 4.71 \\
Fe\,\textsc{ii}  & 4576.34 & 2.84 & $-$2.95 & 8.4 & 4.71 \\
Fe\,\textsc{ii}  & 4582.84 & 2.84 & $-$3.18 & 4.6 & 4.67 \\
Fe\,\textsc{ii}  & 4583.84 & 2.81 & $-$1.93 & 60.6 & 4.77 \\
Fe\,\textsc{ii}  & 4620.52 & 2.83 & $-$3.21 & 3.5 & 4.57 \\
Fe\,\textsc{ii}  & 4923.93 & 2.89 & $-$1.26 & 91.7 & 4.60 \\
Fe\,\textsc{ii}  & 5018.45 & 2.89 & $-$1.10 & 102.9 & 4.61 \\
Fe\,\textsc{ii}  & 5169.03 & 2.89 & $-$1.00 & 122.8 & 4.85 \\
Fe\,\textsc{ii}  & 5197.58 & 3.23 & $-$2.22 & 16.4 & 4.65 \\
Fe\,\textsc{ii}  & 5234.63 & 3.22 & $-$2.18 & 24.5 & 4.81 \\
Fe\,\textsc{ii}  & 5276.00 & 3.20 & $-$2.01 & 26.7 & 4.67 \\
Fe\,\textsc{ii}  & 5316.62 & 3.15 & $-$1.87 & 45.1 & 4.79 \\
Fe\,\textsc{ii}  & 5534.83 & 3.25 & $-$2.75 & 4.1 & 4.54 \\
Co\,\textsc{i} & 3873.12 & 0.43 & $-$0.66 & syn & 2.32 \\
Co\,\textsc{i} & 3995.31 & 0.92 & $-$0.22 & syn & 2.29 \\
Co\,\textsc{i} & 4020.83 & 3.66 & $-$0.96 & syn & 2.43 \\
Co\,\textsc{i} & 4121.31 & 0.92 & $-$0.32 & syn & 2.23 \\
Ni\,\textsc{i} & 3423.71 & 0.21 & $-$0.71 & 73.6 & 3.23 \\
Ni\,\textsc{i} & 3433.56 & 0.03 & $-$0.67 & 97.5 & 3.55 \\
Ni\,\textsc{i} & 3452.89 & 0.11 & $-$0.90 & 88.0 & 3.63 \\
Ni\,\textsc{i} & 3472.54 & 0.11 & $-$0.79 & 76.5 & 3.26 \\
Ni\,\textsc{i} & 3483.78 & 0.28 & $-$1.11 & 48.7 & 3.23 \\
Ni\,\textsc{i} & 3492.96 & 0.11 & $-$0.27 & 94.7 & 3.15 \\
Ni\,\textsc{i} & 3500.85 & 0.17 & $-$1.27 & 53.9 & 3.36 \\
Ni\,\textsc{i} & 3519.76 & 0.28 & $-$1.44 & 51.0 & 3.59 \\
Ni\,\textsc{i} & 3566.37 & 0.42 & $-$0.25 & 84.3 & 3.18 \\
Ni\,\textsc{i} & 3597.70 & 0.21 & $-$1.10 & 59.6 & 3.33 \\
Ni\,\textsc{i} & 3783.53 & 0.42 & $-$1.40 & 57.6 & 3.55 \\
Ni\,\textsc{i} & 3807.14 & 0.42 & $-$1.23 & 63.6 & 3.46 \\
Ni\,\textsc{i} & 3858.30 & 0.42 & $-$0.96 & 81.8 & 3.46 \\
Ni\,\textsc{i} & 5476.90 & 1.82 & $-$0.78 & 21.8 & 3.58 \\
Zn\,\textsc{i} & 4810.53 & 4.08 & $-$0.15 & 2.7 & 1.80 \\
Sr\,\textsc{ii} & 4161.79 & $−$0.60 & 2.94 & syn & 1.24 \\
Y\,\textsc{ii}  & 3611.04 & 0.13 &  0.11  & syn & 0.53 \\
Y\,\textsc{ii}  & 3710.29 &  0.18 & 0.46 & syn & 0.66 \\
Y\,\textsc{ii}  & 4398.01 & 0.13 & $-$1.00 & syn & 0.45 \\
Y\,\textsc{ii}  & 4682.73 & 0.41&  $-$1.51 & syn & 0.42 \\
Y\,\textsc{ii}  & 4854.87 & 0.99& $-$0.38 & syn & 0.45 \\
Y\,\textsc{ii}  & 4883.68 & 1.08&  0.07 & syn & 0.53 \\
Y\,\textsc{ii}  & 4900.11 & 1.03  &$-$0.09 & syn & 0.35 \\
Y\,\textsc{ii}  & 5087.42 & 1.08& $-$0.17 & syn & 0.42 \\
Y\,\textsc{ii}  & 5199.11 & 0.99& $-$1.36 & syn & 0.26 \\
Y\,\textsc{ii}  & 5205.73 & 1.03 & $-$1.34 & syn & 0.39 \\
Zr\,\textsc{ii}  & 3479.02 & 0.53 & $-$0.69 & syn & 0.88 \\
Zr\,\textsc{ii}  & 3499.57 & 0.41 & $-$0.81 & syn & 1.02 \\
Zr\,\textsc{ii}  & 4050.33 & 0.71 & $-$1.00 & syn & 1.14 \\
Zr\,\textsc{ii}  & 4149.20 & 0.80& $-$0.03 & syn & 1.21 \\
Zr\,\textsc{ii}  & 4161.21 & 0.71 &  $-$0.72 & syn & 1.32 \\
Zr\,\textsc{ii}  & 4208.99 & 0.71  &$-$0.46 & syn & 1.09 \\
Zr\,\textsc{ii}  & 4317.32 &0.71 &$-$1.38  & syn & 1.18 \\
Zr\,\textsc{ii}  & 4613.95 & 0.97 & $-$1.52  & syn & 1.22 \\
Ru\,\textsc{i} & 3728.03 & 0.00 & 0.27 & syn & 0.89\\
Ba\,\textsc{ii} & 4130.65 & 2.72 & 0.68 & syn & 0.63 \\
Ba\,\textsc{ii} & 5853.69 & 0.60 & $-$0.91 & syn & 0.84 \\
La\,\textsc{ii} & 3794.77 & 0.24 & 0.21 & syn & 0.00 \\
La\,\textsc{ii} & 3849.01 & 0.00 & $-$0.45 & syn & 0.13 \\
La\,\textsc{ii} & 3929.21 & 0.17 & $-$0.32 & syn & 0.02 \\
La\,\textsc{ii} & 3949.10 & 0.00 & 0.49 & syn & 0.04 \\
La\,\textsc{ii} & 3988.51 & 0.40 & 0.21 & syn & 0.10 \\
La\,\textsc{ii} & 4086.71 & 0.00 & $-$0.07  & syn & 0.19 \\
La\,\textsc{ii} & 4123.22 & 0.32 & 0.13 & syn & 0.09 \\
La\,\textsc{ii} & 4141.72 & 0.40 & $-$0.66 & syn & 0.06 \\
La\,\textsc{ii} & 4322.51 & 0.17 & $-$0.93  & syn & 0.17 \\
La\,\textsc{ii} & 4333.75 & 0.17 & $-$0.06  & syn & 0.10 \\
La\,\textsc{ii} & 4429.91 & 0.23 & $-$0.35 & syn & 0.09 \\
La\,\textsc{ii} & 4526.12 & 0.77 & $-$0.59 & syn & 0.05 \\
La\,\textsc{ii} & 4574.88 & 0.17 & $-$1.08 & syn & 0.08 \\
La\,\textsc{ii} & 4662.51 & 0.00 & $-$1.24 & syn & 0.22 \\
La\,\textsc{ii} & 4748.73 & 0.93 & $-$0.54 & syn & 0.12 \\
La\,\textsc{ii} & 4921.78 & 0.24 & $-$0.45 & syn & 0.15 \\
La\,\textsc{ii} & 5114.56 & 0.23 & $-$1.03 & syn & 0.25 \\
La\,\textsc{ii} & 5122.99 & 0.32 & $−$0.85 & syn & 0.19 \\
Ce\,\textsc{ii} & 3942.15 & 0.00 & $-$0.22 & 31.4 & 0.58 \\
Ce\,\textsc{ii} & 3942.74 & 0.86 & 0.69 & 26.4 & 0.42 \\
Ce\,\textsc{ii} & 4083.22 & 0.70 & 0.27 & 17.2 & 0.45 \\
Ce\,\textsc{ii} & 4120.83 & 0.32 & $-$0.37 & 10.3 & 0.46 \\
Ce\,\textsc{ii} & 4127.36 & 0.68 & 0.31 & 16.4 & 0.36 \\
Ce\,\textsc{ii} & 4137.65 & 0.52 & 0.40 & 32.4 & 0.47 \\
Ce\,\textsc{ii} & 4142.40 & 0.70 & 0.22 & 14.1 & 0.39 \\
Ce\,\textsc{ii} & 4145.00 & 0.70 & 0.10 & 12.1 & 0.43 \\
Ce\,\textsc{ii} & 4222.60 & 0.12 & $-$0.15 & 23.3 & 0.43 \\
Ce\,\textsc{ii} & 4364.65 & 0.49 & $-$0.17 & 7.7 & 0.27 \\
Ce\,\textsc{ii} & 4382.16 & 0.68 & 0.13 & 11.5 & 0.34 \\
Ce\,\textsc{ii} & 4399.20 & 0.33 & $-$0.44 & 6.0 & 0.25 \\
Ce\,\textsc{ii} & 4449.33 & 0.61 & 0.04 & 13.2 & 0.41 \\
Ce\,\textsc{ii} & 4486.91 & 0.29 & $-$0.18 & 14.0 & 0.35 \\
Ce\,\textsc{ii} & 4523.07 & 0.52 & $-$0.08 & 22.6 & 0.71 \\
Ce\,\textsc{ii} & 4560.28 & 0.91 & 0.18 & 12.4 & 0.54 \\
Ce\,\textsc{ii} & 4560.96 & 0.68 & $-$0.26 & 6.1 & 0.42 \\
Ce\,\textsc{ii} & 4562.36 & 0.48 & 0.21 & 23.4 & 0.39 \\
Ce\,\textsc{ii} & 4593.93 & 0.70 & 0.07 & 13.3 & 0.47 \\
Ce\,\textsc{ii} & 4628.16 & 0.52 & 0.14 & 19.7 & 0.41 \\
Pr\,\textsc{ii} & 4062.81 & 0.42 & 0.33 & syn & $-$0.06 \\
Pr\,\textsc{ii} & 4143.13 & 0.37 & 0.60 & syn & $-$0.16 \\
Pr\,\textsc{ii} & 4164.16 & 0.20 & 0.17 & syn & $-$0.25 \\
Pr\,\textsc{ii} & 4179.39 & 0.20 & 0.48 & syn & $-$0.21 \\
Pr\,\textsc{ii} & 4189.48 & 0.37 & 0.38 & syn & $-$0.18 \\
Pr\,\textsc{ii} & 4222.95 & 0.06 & 0.27 & syn & $-$0.17 \\
Pr\,\textsc{ii} & 4408.82 & 0.00 & 0.18 & syn & $-$0.22 \\
Pr\,\textsc{ii} & 4449.83 & 0.20 & $-$0.26 & syn & 0.02 \\
Nd\,\textsc{ii} & 3784.24 & 0.38 & 0.15 & 28.8 & 0.38 \\
Nd\,\textsc{ii} & 3927.10 & 0.18 & $-$0.59 & 10.9 & 0.41 \\
Nd\,\textsc{ii} & 3990.10 & 0.47 & 0.13 & 31.0 & 0.51 \\
Nd\,\textsc{ii} & 4007.43 & 0.47 & $-$0.40 & 9.8 & 0.45 \\
Nd\,\textsc{ii} & 4012.70 & 0.00 & $-$0.60 & 18.5 & 0.49 \\
Nd\,\textsc{ii} & 4021.33 & 0.32 & $-$0.10 & 20.5 & 0.36 \\
Nd\,\textsc{ii} & 4023.00 & 0.56 & 0.04 & 27.2 & 0.61 \\
Nd\,\textsc{ii} & 4043.59 & 0.32 & $-$0.71 & 6.2 & 0.39 \\
Nd\,\textsc{ii} & 4051.14 & 0.38 & $-$0.30 & 14.6 & 0.44 \\
Nd\,\textsc{ii} & 4059.95 & 0.20 & $-$0.52 & 10.3 & 0.32 \\
Nd\,\textsc{ii} & 4061.08 & 0.47 & 0.55 & 54.7 & 0.46 \\
Nd\,\textsc{ii} & 4069.26 & 0.06 & $-$0.57 & 15.5 & 0.42 \\
Nd\,\textsc{ii} & 4109.45 & 0.32 & 0.35 & 50.6 & 0.44 \\
Nd\,\textsc{ii} & 4133.35 & 0.32 & $-$0.49 & 10.5 & 0.41 \\
Nd\,\textsc{ii} & 4135.32 & 0.63 & $-$0.07 & 21.0 & 0.64 \\
Nd\,\textsc{ii} & 4211.29 & 0.20 & $-$0.86 & 6.1 & 0.40 \\
Nd\,\textsc{ii} & 4232.37 & 0.06 & $-$0.47 & 20.7 & 0.46 \\
Nd\,\textsc{ii} & 4284.51 & 0.63 & $-$0.17 & 11.7 & 0.43 \\
Nd\,\textsc{ii} & 4351.28 & 0.18 & $-$0.61 & 15.0 & 0.54 \\
Nd\,\textsc{ii} & 4358.16 & 0.32 & $-$0.16 & 21.0 & 0.39 \\
Nd\,\textsc{ii} & 4368.63 & 0.06 & $-$0.81 & 9.2 & 0.39 \\
Nd\,\textsc{ii} & 4385.66 & 0.20 & $-$0.30 & 22.5 & 0.45 \\
Nd\,\textsc{ii} & 4400.82 & 0.06 & $-$0.60 & 10.1 & 0.21 \\
Nd\,\textsc{ii} & 4446.38 & 0.20 & $-$0.35 & 18.1 & 0.38 \\
Nd\,\textsc{ii} & 4462.98 & 0.56 & 0.04 & 22.0 & 0.45 \\
Nd\,\textsc{ii} & 4465.06 & 0.00 & $-$1.36 & 4.2 & 0.51 \\
Nd\,\textsc{ii} & 4501.81 & 0.20 & $-$0.69 & 9.8 & 0.42 \\
Nd\,\textsc{ii} & 4542.60 & 0.74 & $-$0.28 & 7.0 & 0.39 \\
Nd\,\textsc{ii} & 4563.22 & 0.18 & $-$0.88 & 10.3 & 0.61 \\
Nd\,\textsc{ii} & 4567.61 & 0.20 & $-$1.31 & 3.7 & 0.59 \\
Nd\,\textsc{ii} & 4645.76 & 0.56 & $-$0.76 & 3.7 & 0.39 \\
Nd\,\textsc{ii} & 4706.54 & 0.00 & $-$0.71 & 14.0 & 0.39 \\
Nd\,\textsc{ii} & 4709.72 & 0.18 & $-$0.97 & 7.6 & 0.54 \\
Nd\,\textsc{ii} & 4715.59 & 0.20 & $-$0.90 & 5.3 & 0.33 \\
Nd\,\textsc{ii} & 4820.34 & 0.20 & $-$0.92 & 8.3 & 0.55 \\
Nd\,\textsc{ii} & 4825.48 & 0.18 & $-$0.42 & 16.6 & 0.35 \\
Nd\,\textsc{ii} & 4914.38 & 0.38 & $-$0.70 & 6.4 & 0.38 \\
Nd\,\textsc{ii} & 4959.12 & 0.06 & $-$0.80 & 10.3 & 0.38 \\
Nd\,\textsc{ii} & 5092.79 & 0.38 & $-$0.61 & 14.1 & 0.64 \\
Nd\,\textsc{ii} & 5130.59 & 1.30 & 0.45 & 14.5 & 0.52 \\
Nd\,\textsc{ii} & 5234.19 & 0.55 & $-$0.51 & 10.6 & 0.57 \\
Nd\,\textsc{ii} & 5249.58 & 0.97 & 0.20 & 12.6 & 0.37 \\
Nd\,\textsc{ii} & 5255.51 & 0.20 & $-$0.67 & 9.8 & 0.34 \\
Nd\,\textsc{ii} & 5273.43 & 0.68 & $-$0.18 & 12.8 & 0.46 \\
Nd\,\textsc{ii} & 5293.16 & 0.82 & 0.10 & 17.2 & 0.47 \\
Nd\,\textsc{ii} & 5319.81 & 0.55 & $-$0.14 & 22.7 & 0.57 \\
Sm\,\textsc{ii} & 3568.89 & 0.00 &$-2.15$ & syn & 0.11 \\
Sm\,\textsc{ii} & 4065.01 & 0.04 &$-$2.34 & syn & $-$0.07 \\
Sm\,\textsc{ii} & 4511.83 & 0.18 & $-0.82$ & syn & 0.18 \\
Sm\,\textsc{ii} & 4519.63 & 0.54 & $-$0.35 & syn & 0.10 \\
Sm\,\textsc{ii} & 4537.94 & 0.49 & $-$0.48 & syn & 0.10 \\
Sm\,\textsc{ii} & 4554.44 & 0.10 & $-$1.25 & syn & 0.14 \\
Sm\,\textsc{ii} & 4591.81 & 0.18 & $-$1.12 & syn & 0.13 \\
Sm\,\textsc{ii} & 4595.28 & 0.49 & $-$0.50 & syn & 0.12 \\
Sm\,\textsc{ii} & 4595.28 & 0.49 & $-$0.50 & syn & 0.12 \\
Sm\,\textsc{ii} & 4605.17 & 0.04 & $-$1.39 & syn & $-$0.02 \\
Sm\,\textsc{ii} & 4686.19 & 0.04 & $-$1.15 & syn & 0.15 \\
Sm\,\textsc{ii} & 4719.84 & 0.04 & $-$1.24 & syn & 0.05 \\
Sm\,\textsc{ii} & 4745.68 & 0.10&  $-$0.93 & syn & 0.14 \\
Eu\,\textsc{ii} & 3724.93 & 0.00 & $-$0.09 & syn & $-$0.11 \\
Eu\,\textsc{ii} & 3819.67 & 0.00 & 0.51 & syn & $-$0.04 \\
Eu\,\textsc{ii} & 3907.11 & 0.21 & 0.17 & syn & $-$0.12 \\
Eu\,\textsc{ii} & 4129.72 & 0.00 & 0.22 & syn & $-$0.11 \\
Eu\,\textsc{ii} & 4205.04 & 0.00 & 0.21 & syn & $-$0.16 \\
Eu\,\textsc{ii} & 6645.06 & 1.38 & 0.12 & syn & 0.04 \\
Eu\,\textsc{ii} & 7217.56 & 1.23 & $-$0.35 & syn & 0.22 \\
Gd\,\textsc{ii} & 3768.40 & 0.08 & 0.21 & 51.3 & 0.22 \\
Gd\,\textsc{ii} & 3894.69 & 0.00 & $-$0.58 & 29.6 & 0.56 \\
Gd\,\textsc{ii} & 4037.32 & 0.66 & $-$0.11 & 14.8 & 0.36 \\
Gd\,\textsc{ii} & 4049.42 & 0.66 & $-$0.08 & 14.6 & 0.32 \\
Gd\,\textsc{ii} & 4049.85 & 0.99 & 0.49 & 30.5 & 0.47 \\
Gd\,\textsc{ii} & 4191.07 & 0.43 & $-$0.48 & 14.4 & 0.47 \\
Gd\,\textsc{ii} & 4251.73 & 0.38 & $-$0.22 & 24.1 & 0.42 \\
Gd\,\textsc{ii} & 4316.05 & 0.66 & $-$0.45 & 12.3 & 0.58 \\
Gd\,\textsc{ii} & 4438.25 & 0.66 & $-$0.82 & 6.3 & 0.63 \\
Tb\,\textsc{ii} & 3568.45 & 0.00 & 0.36 & syn & $-$0.42 \\
Tb\,\textsc{ii} & 3658.89 & 0.13 & $-$0.01 & syn & $-$0.20 \\
Tb\,\textsc{ii} & 3899.19 & 0.37 & 0.33 & syn & $-$0.47 \\
Dy\,\textsc{ii} & 3996.69 & 0.59 & $-$0.26 & syn & 0.43 \\
Dy\,\textsc{ii} & 4050.57 & 0.59 & $-$0.47 & syn & 0.45 \\
Dy\,\textsc{ii} & 4073.12 & 0.54 & $-$0.32 & syn & 0.43 \\
Dy\,\textsc{ii} & 4077.97 & 0.10 & $-$0.04 & syn & 0.45 \\
Dy\,\textsc{ii} & 4103.31 & 0.10 & $-$0.38 & syn & 0.48 \\
Ho\,\textsc{ii} & 3398.94 & 0.00 & 0.41 & syn & $-$0.19 \\
Ho\,\textsc{ii} & 3416.44 & 0.08 & 0.26 & syn & $-$0.46 \\
Ho\,\textsc{ii} & 3453.11 & 0.08 & 0.01 & syn & $-$0.24 \\
Ho\,\textsc{ii} & 3456.01 & 0.00 & 0.76 & syn & $-$0.21 \\
Ho\,\textsc{ii} & 3474.27 & 0.08 & 0.28 & syn & $-$0.23 \\
Ho\,\textsc{ii} & 3484.83 & 0.08 & 0.28 & syn & $-$0.35 \\
Ho\,\textsc{ii} & 3810.71 & 0.00 & 0.19 & syn & $-$0.33 \\
Ho\,\textsc{ii} & 3890.97 & 0.08 & 0.46 & syn & $-$0.41 \\
Ho\,\textsc{ii} & 4045.45 & 0.00 & $-$0.05 & syn & $-$0.27 \\
Er\,\textsc{ii} & 3499.11 & 0.06 & 0.29 & 57.1 & 0.14 \\
Er\,\textsc{ii} & 3559.89 & 0.00 & $-$0.69 & 17.4 & 0.26 \\
Er\,\textsc{ii} & 3616.57 & 0.00 & $-$0.31 & 26.3 & 0.10 \\
Er\,\textsc{ii} & 3633.54 & 0.00 & $-$0.53 & 28.3 & 0.36 \\
Er\,\textsc{ii} & 3896.23 & 0.06 & $-$0.12 & 49.6 & 0.18 \\
Tm\,\textsc{ii} & 3462.20 & 0.00 & 0.03 & syn & $-$0.71 \\
Tm\,\textsc{ii} & 3761.91 & 0.00 & $-$0.43  & syn & $-$0.83 \\
Tm\,\textsc{ii} & 3795.17 & 0.03 & $-$1.58 & syn & $-$0.45 \\
Tm\,\textsc{ii} & 3848.02 & 0.00 & $-$0.14 & syn & $-$0.65 \\
Tm\,\textsc{ii} & 3996.51 & 0.00 & $-$1.20 & syn & $-$0.48 \\
Yb\,\textsc{ii} & 3694.19 & 0.00 & $-$0.30 & syn & 0.02 \\
Hf\,\textsc{ii} & 4093.15 & 0.45 & $-$1.15 & syn & $-$0.03 \\
Os\,\textsc{i} & 4420.46 & 0.00 & $-$1.43 & syn & 0.90 \\
Ir\,\textsc{i} & 3513.65 & 0.00 & $-$1.26  & syn & $<$1.53 \\
Th\,\textsc{ii} & 4019.12 & 0.00 & $-$0.23  & syn & $-$0.60 \\
U\,\textsc{ii} & 3859.57 & 0.04 & $-$0.07 & syn & $<-$0.59 \\
\enddata
\tablecomments{``syn" denotes spectrum synthesis was used to measure the abundance.}
\end{deluxetable}
\(\)

\begin{thebibliography}{}
\expandafter\ifx\csname natexlab\endcsname\relax\def\natexlab#1{#1}\fi

\bibitem[{{Abbott} {et~al.}(2017{\natexlab{a}}){Abbott}, {Abbott}, {Abbott},
  {Acernese}, {Ackley}, {Adams}, {Adams}, {Addesso}, {Adhikari}, {Adya}, \&
  et~al.}]{LIGOGW170817a}
{Abbott}, B.~P., {Abbott}, R., {Abbott}, T.~D., {et~al.} 2017{\natexlab{a}},
  Physical Review Letters, 119, 161101

\bibitem[{{Abbott} {et~al.}(2017{\natexlab{b}}){Abbott}, {Abbott}, {Abbott},
  {Acernese}, {Ackley}, {Adams}, {Adams}, {Addesso}, {Adhikari}, {Adya}, \&
  et~al.}]{LIGOGW170817b}
---. 2017{\natexlab{b}}, \apjl, 848, L12

\bibitem[{{Abohalima} \& {Frebel}(2018)}]{Abohalima18}
{Abohalima}, A., \& {Frebel}, A. 2018, \apjs, 238, 36

\bibitem[{{Aoki} {et~al.}(2007){Aoki}, {Beers}, {Christlieb}, {Norris}, {Ryan},
  \& {Tsangarides}}]{Aoki07}
{Aoki}, W., {Beers}, T.~C., {Christlieb}, N., {et~al.} 2007, \apj, 655, 492

\bibitem[{{Aoki} {et~al.}(2002{\natexlab{a}}){Aoki}, {Norris}, {Ryan}, {Beers},
  \& {Ando}}]{aoki02}
{Aoki}, W., {Norris}, J.~E., {Ryan}, S.~G., {Beers}, T.~C., \& {Ando}, H.
  2002{\natexlab{a}}, \pasj, 54, 933

\bibitem[{{Aoki} {et~al.}(2002{\natexlab{b}}){Aoki}, {Ando}, {Honda}, {Iye},
  {Izumiura}, {Kajino}, {Kambe}, {Kawanomonoto}, {Noguchi}, {Okita},
  {Sadakane}, {Sato}, {Shelton}, {Takada-Hidai}, {Takeda}, {Watanabe}, \&
  {Yoshida}}]{aoki2002_lp625}
{Aoki}, W., {Ando}, H., {Honda}, S., {et~al.} 2002{\natexlab{b}}, PASJ, 54, 427

\bibitem[{{Arcones} \& {Montes}(2011)}]{Arcones11}
{Arcones}, A., \& {Montes}, F. 2011, \apj, 731, 5

\bibitem[{{Arlandini} {et~al.}(1999){Arlandini}, {K\"{a}ppeler}, {Wisshak}, {Gallino}, {Lugaro}, {Busso}, \& {Straniero}}]{Arlandini99}
{Arlandini}, C., {K\"{a}ppeler}, F., {Wisshak}, K., {et~al.} 1999, \apj, 525, 2

\bibitem[{{Asplund} {et~al.}(2009){Asplund}, {Grevesse}, {Sauval}, \&
  {Scott}}]{Asplund09}
{Asplund}, M., {Grevesse}, N., {Sauval}, A.~J., \& {Scott}, P. 2009, \araa, 47,
  481

\bibitem[{{Astropy Collaboration} {et~al.}(2013){Astropy Collaboration},
  {Robitaille}, {Tollerud}, {Greenfield}, {Droettboom}, {Bray}, {Aldcroft},
  {Davis}, {Ginsburg}, {Price-Whelan}, {Kerzendorf}, {Conley}, {Crighton},
  {Barbary}, {Muna}, {Ferguson}, {Grollier}, {Parikh}, {Nair}, {Unther},
  {Deil}, {Woillez}, {Conseil}, {Kramer}, {Turner}, {Singer}, {Fox}, {Weaver},
  {Zabalza}, {Edwards}, {Azalee Bostroem}, {Burke}, {Casey}, {Crawford},
  {Dencheva}, {Ely}, {Jenness}, {Labrie}, {Lim}, {Pierfederici}, {Pontzen},
  {Ptak}, {Refsdal}, {Servillat}, \& {Streicher}}]{astropy}
{Astropy Collaboration}, {Robitaille}, T.~P., {Tollerud}, E.~J., {et~al.} 2013,
  \aap, 558, A33

\bibitem[{{Barbuy} {et~al.}(2005){Barbuy}, {Spite}, {Spite}, {Hill}, {Cayrel},
  {Plez}, \& {Petitjean}}]{barbuy2005}
{Barbuy}, B., {Spite}, M., {Spite}, F., {et~al.} 2005, \aap, 429, 1031

\bibitem[{{Barklem}(2018)}]{Barklem2018}
{Barklem}, P.~S. 2018, \aap, 612, A90

\bibitem[{{Barklem} {et~al.}(2005){Barklem}, {Christlieb}, {Beers}, {Hill},
  {Bessell}, {Holmberg}, {Marsteller}, {Rossi}, {Zickgraf}, \&
  {Reimers}}]{Barklem05}
{Barklem}, P.~S., {Christlieb}, N., {Beers}, T.~C., {et~al.} 2005, \aap, 439,
  129

\bibitem[{{Beers} \& {Christlieb}(2005)}]{Beers05}
{Beers}, T.~C., \& {Christlieb}, N. 2005, \araa, 43, 531

\bibitem[{{Beers} {et~al.}(2014){Beers}, {Norris}, {Placco}, {Lee}, {Rossi},
  {Carollo}, \& {Masseron}}]{beers14}
{Beers}, T.~C., {Norris}, J.~E., {Placco}, V.~M., {et~al.} 2014, \apj, 794, 58

\bibitem[{{Beers} {et~al.}(2017){Beers}, {Placco}, {Carollo}, {Rossi}, {Lee},
  {Frebel}, {Norris}, {Dietz}, \& {Masseron}}]{beers17}
{Beers}, T.~C., {Placco}, V.~M., {Carollo}, D., {et~al.} 2017, \apj, 835, 81

\bibitem[{{Bernstein} {et~al.}(2003){Bernstein}, {Shectman}, {Gunnels},
  {Mochnacki}, \& {Athey}}]{Bernstein03}
{Bernstein}, R., {Shectman}, S.~A., {Gunnels}, S.~M., {Mochnacki}, S., \&
  {Athey}, A.~E. 2003, Proc. SPIE, 4841, 1694

\bibitem[{{Binney}(2012)}]{binney12}
{Binney}, J. 2012, \mnras, 426, 1324

\bibitem[{{Bovy}(2015)}]{bovy15}
{Bovy}, J. 2015, \apjs, 216, 29

\bibitem[{{Brauer} {et~al.}(2019){Brauer}, {Ji}, {Frebel}, {Dooley},
  {G{\'o}mez}, \& {O'Shea}}]{brauer19}
{Brauer}, K., {Ji}, A.~P., {Frebel}, A., {et~al.} 2019, \apj, 871, 247

\bibitem[{{Burris} {et~al.}(2000){Burris}, {Pilachowski}, {Armandroff},
  {Sneden}, {Cowan}, \& {Roe}}]{Burris00}
{Burris}, D.~L., {Pilachowski}, C.~A., {Armandroff}, T.~E., {et~al.} 2000,
  \apj, 544, 302

\bibitem[{{Cain} {et~al.}(2018){Cain}, {Frebel}, {Gull}, {Ji}, {Placco},
  {Beers}, {Mel{\'e}ndez}, {Ezzeddine}, {Casey}, {Hansen}, {Roederer}, \&
  {Sakari}}]{cain18}
{Cain}, M., {Frebel}, A., {Gull}, M., {et~al.} 2018, \apj, 864, 43

\bibitem[{{Casey}(2014)}]{Casey14}
{Casey}, A.~R. 2014, ArXiv e-prints, arXiv:1405.5968

\bibitem[{{Casey} \& {Schlaufman}(2017)}]{casey17}
{Casey}, A.~R., \& {Schlaufman}, K.~C. 2017, \apj, 850, 179

\bibitem[{{Casey} {et~al.}(2017){Casey}, {Hawkins}, {Hogg}, {Ness}, {Rix}, {Kordopatis}, 
{Kunder}, {Steinmetz}, {Koposov}, {Enke}, {Sanders}, {Gilmore}, {Zwitter}, {Freeman}, {Casagrande}, {Matijevi{\v{c}}}, {Seabroke},
 {Bienaym{\'e}}, {Bland-Hawthorn}, {Gibson}, {Grebel}, {Helmi}, {Munari}, {Navarro}, {Reid},
 {Siebert}, {Wyse}}]{casey2017} {Casey}, A.~R., {Hawkins}, K., {Hogg}, D.~W., {et al.} 2017, \apj, 840, 59

\bibitem[{{Castelli} \& {Kurucz}(2004)}]{Castelli04}
{Castelli}, F., \& {Kurucz}, R.~L. 2004, ArXiv Astrophysics e-prints,
  astro-ph/0405087

\bibitem[{{Cayrel} {et~al.}(2001){Cayrel}, {Depagne}, {Hill}, {Beers},
   {Barbuy}, {Spite}, {Plez}, {Andersen}, {Bonifacio}, {Fran{\c c}ois}, {Molaro}, {Nordstr{\"o}m}, \& {Primas}}]{Cayreletal:2001}
{Cayrel}, R., {Depagne}, E., {Spite}, M., {et~al.} 2004, \aap, 416, 1117

\bibitem[{{Cayrel} {et~al.}(2004){Cayrel}, {Depagne}, {Spite}, {Hill}, {Spite},
  {Fran{\c c}ois}, {Plez}, {Beers}, {Primas}, {Andersen}, {Barbuy},
  {Bonifacio}, {Molaro}, \& {Nordstr{\"o}m}}]{Cayrel04}
{Cayrel}, R., {Hill}, V., {Beers}, T.C., {et~al.} 2001, Nature, 409, 691-692

\bibitem[{{Cohen} {et~al.}(2003){Cohen}, {Christlieb}, {Qian}, \&
  {Wasserburg}}]{cohen03}
{Cohen}, J.~G., {Christlieb}, N., {Qian}, Y.-Z., \& {Wasserburg}, G.~J. 2003,
  \apj, 588, 1082

\bibitem[{{Cohen} {et~al.}(2006){Cohen}, {McWilliam}, {Shectman}, {Thompson},
  {Christlieb}, {Melendez}, {Ramirez}, {Swensson}, \& {Zickgraf}}]{cohen06}
{Cohen}, J.~G., {McWilliam}, A., {Shectman}, S., {et~al.} 2006, \aj, 132, 137

\bibitem[{{Coulter} {et~al.}(2017){Coulter}, {Foley}, {Kilpatrick}, {Drout},
  {Piro}, {Shappee}, {Siebert}, {Simon}, {Ulloa}, {Kasen}, {Madore},
  {Murguia-Berthier}, {Pan}, {Prochaska}, {Ramirez-Ruiz}, {Rest}, \&
  {Rojas-Bravo}}]{coulter17}
{Coulter}, D.~A., {Foley}, R.~J., {Kilpatrick}, C.~D., {et~al.} 2017, Science,
  358, 1556

\bibitem[{{Cui} {et~al.}(2013){Cui}, {Sivarani}, \& {Christlieb}}]{cui13}
{Cui}, W.~Y., {Sivarani}, T., \& {Christlieb}, N. 2013, \aap, 558, A36

\bibitem[{{Dardelet} {et~al.}(2014){Dardelet}, {Ritter}, {Prado}, {Heringer},
  {Higgs}, {Sandalski}, {Jones}, {Denisenkov}, {Venn}, {Bertolli}, {Pignatari},
  {Woodward}, \& {Herwig}}]{dardelet14}
{Dardelet}, L., {Ritter}, C., {Prado}, P., {et~al.} 2014, in XIII Nuclei in the
  Cosmos (NIC XIII), 145

\bibitem[{{Den Hartog} {et~al.}(2014){Den Hartog}, {Ruffoni}, {Lawler},
  {Pickering}, {Lind}, \& {Brewer}}]{den14}
{Den Hartog}, E.~A., {Ruffoni}, M.~P., {Lawler}, J.~E., {et~al.} 2014, \apjs,
  215, 23

\bibitem[{{Drout} {et~al.}(2017){Drout}, {Piro}, {Shappee}, {Kilpatrick},
  {Simon}, {Contreras}, {Coulter}, {Foley}, {Siebert}, {Morrell}, {Boutsia},
  {Di Mille}, {Holoien}, {Kasen}, {Kollmeier}, {Madore}, {Monson},
  {Murguia-Berthier}, {Pan}, {Prochaska}, {Ramirez-Ruiz}, {Rest}, {Adams},
  {Alatalo}, {Ba{\~n}ados}, {Baughman}, {Beers}, {Bernstein}, {Bitsakis},
  {Campillay}, {Hansen}, {Higgs}, {Ji}, {Maravelias}, {Marshall}, {Bidin},
  {Prieto}, {Rasmussen}, {Rojas-Bravo}, {Strom}, {Ulloa},
  {Vargas-Gonz{\'a}lez}, {Wan}, \& {Whitten}}]{drout17}
{Drout}, M.~R., {Piro}, A.~L., {Shappee}, B.~J., {et~al.} 2017, Science, 358,
  1570

\bibitem[{{Ezzeddine} {et~al.}(2017){Ezzeddine}, {Frebel}, \&
  {Plez}}]{Ezzeddine2017}
{Ezzeddine}, R., {Frebel}, A., \& {Plez}, B. 2017, \apj, 847, 142

\bibitem[{{Ezzeddine} {et~al.}(2016){Ezzeddine}, {Merle}, \&
  {Plez}}]{Ezzeddine2016}
{Ezzeddine}, R., {Merle}, T., \& {Plez}, B. 2016, Astronomische Nachrichten,
  337, 850

\bibitem[{{Ezzeddine} {et~al.}(2018){Ezzeddine}, {Merle}, {Plez}, {Gebran},
  {Th{\'e}venin}, \& {Van der Swaelmen}}]{ezzeddine16a}
{Ezzeddine}, R., {Merle}, T., {Plez}, B., {et~al.} 2018, \aap, 618, A141

\bibitem[{{Frebel}(2018)}]{frebel18}
{Frebel}, A. 2018, Annual Review of Nuclear and Particle Science, 68, 237

\bibitem[{{Frebel} {et~al.}(2013){Frebel}, {Casey}, {Jacobson}, \&
  {Yu}}]{frebel13a}
{Frebel}, A., {Casey}, A.~R., {Jacobson}, H.~R., \& {Yu}, Q. 2013, ApJ, 769, 57

\bibitem[{{Frebel} {et~al.}(2007){Frebel}, {Christlieb}, {Norris}, {Thom},
  {Beers}, \& {Rhee}}]{Frebel07}
{Frebel}, A., {Christlieb}, N., {Norris}, J.~E., {et~al.} 2007, \apjl, 660,
  L117

\bibitem[{{Frebel} \& {Norris}(2015)}]{Frebel15}
{Frebel}, A., \& {Norris}, J.~E. 2015, \araa, 53, 631

\bibitem[{{Gaia Collaboration} {et~al.}(2016{\natexlab{a}}){Gaia
  Collaboration}, {Brown}, {Vallenari}, {Prusti}, {de Bruijne}, {Mignard},
  {Drimmel}, {Babusiaux}, {Bailer-Jones}, {Bastian}, \& et~al.}]{gaiab}
{Gaia Collaboration}, {Brown}, A.~G.~A., {Vallenari}, A., {et~al.}
  2016{\natexlab{a}}, \aap, 595, A2

\bibitem[{{Gaia Collaboration} {et~al.}(2016{\natexlab{b}}){Gaia
  Collaboration}, {Prusti}, {de Bruijne}, {Brown}, {Vallenari}, {Babusiaux},
  {Bailer-Jones}, {Bastian}, {Biermann}, {Evans}, \& et~al.}]{gaiaa}
{Gaia Collaboration}, {Prusti}, T., {de Bruijne}, J.~H.~J., {et~al.}
  2016{\natexlab{b}}, \aap, 595, A1

\bibitem[{{Gull} {et~al.}(2018){Gull}, {Frebel}, {Cain}, {Placco}, {Ji},
  {Abate}, {Ezzeddine}, {Karakas}, {Hansen}, {Sakari}, {Holmbeck}, {Santucci},
  {Casey}, \& {Beers}}]{gull18}
{Gull}, M., {Frebel}, A., {Cain}, M.~G., {et~al.} 2018, \apj, 862, 174

\bibitem[{{Hampel} {et~al.}(2016){Hampel}, {Stancliffe}, {Lugaro}, \&
  {Meyer}}]{hampel16}
{Hampel}, M., {Stancliffe}, R.~J., {Lugaro}, M., \& {Meyer}, B.~S. 2016, \apj,
  831, 171

\bibitem[{{Hansen} {et~al.}(2015{\natexlab{a}}){Hansen}, {Hansen},
  {Christlieb}, {Beers}, {Yong}, {Bessell}, {Frebel}, {Garc{\'\i}a P{\'e}rez},
  {Placco}, {Norris}, \& {Asplund}}]{hansen15a}
{Hansen}, T., {Hansen}, C.~J., {Christlieb}, N., {et~al.} 2015{\natexlab{a}},
  \apj, 807, 173

\bibitem[{{Hansen} {et~al.}(2015{\natexlab{b}}){Hansen}, {Andersen},
  {Nordstr{\"o}m}, {Beers}, {Yoon}, \& {Buchhave}}]{Hansen15}
{Hansen}, T.~T., {Andersen}, J., {Nordstr{\"o}m}, B., {et~al.}
  2015{\natexlab{b}}, \aap, 583, A49

\bibitem[{{Hansen} {et~al.}(2018){Hansen}, {Holmbeck}, {Beers}, {Placco},
  {Roederer}, {Frebel}, {Sakari}, {Simon}, \& {Thompson}}]{hansen18}
{Hansen}, T.~T., {Holmbeck}, E.~M., {Beers}, T.~C., {et~al.} 2018, \apj, 858,
  92

\bibitem[{{Hayek} {et~al.}(2009){Hayek}, {Wiesendahl}, {Christlieb},
  {Eriksson}, {Korn}, {Barklem}, {Hill}, {Beers}, {Farouqi}, {Pfeiffer}, \&
  {Kratz}}]{hayek09}
{Hayek}, W., {Wiesendahl}, U., {Christlieb}, N., {et~al.} 2009, \aap, 504, 511

\bibitem[{{Hill} {et~al.}(2017){Hill}, {Christlieb}, {Beers}, {Barklem},
  {Kratz}, {Nordstr{\"o}m}, {Pfeiffer}, \& {Farouqi}}]{hill17}
{Hill}, V., {Christlieb}, N., {Beers}, T.~C., {et~al.} 2017, \aap, 607, A91

\bibitem[{{Hill} {et~al.}(2002){Hill}, {Plez}, {Cayrel}, {Beers},
  {Nordstr{\"o}m}, {Andersen}, {Spite}, {Spite}, {Barbuy}, {Bonifacio},
  {Depagne}, {Fran{\c c}ois}, \& {Primas}}]{Hill02}
{Hill}, V., {Plez}, B., {Cayrel}, R., {et~al.} 2002, \aap, 387, 560

\bibitem[{{Hollek} {et~al.}(2015){Hollek}, {Frebel}, {Placco}, {Karakas},
  {Shetrone}, {Sneden}, \& {Christlieb}}]{hollek15}
{Hollek}, J.~K., {Frebel}, A., {Placco}, V.~M., {et~al.} 2015, \apj, 814, 121

\bibitem[{{Holmbeck} {et~al.}(2018){Holmbeck}, {Beers}, {Roederer}, {Placco},
  {Hansen}, {Sakari}, {Sneden}, {Liu}, {Lee}, {Cowan}, \&
  {Frebel}}]{holmbeck18}
{Holmbeck}, E.~M., {Beers}, T.~C., {Roederer}, I.~U., {et~al.} 2018, \apjl,
  859, L24

\bibitem[{{Honda} {et~al.}(2011){Honda}, {Aoki}, {Beers}, \&
  {Takada-Hidai}}]{honda11}
{Honda}, S., {Aoki}, W., {Beers}, T.~C., \& {Takada-Hidai}, M. 2011, \apj, 730,
  77

\bibitem[{{Honda} {et~al.}(2004){Honda}, {Aoki}, {Kajino}, {Ando}, {Beers},
  {Izumiura}, {Sadakane}, \& {Takada-Hidai}}]{Honda04}
{Honda}, S., {Aoki}, W., {Kajino}, T., {et~al.} 2004, \apj, 607, 474

\bibitem[{Hunter(2007)}]{matplotlib}
Hunter, J.~D. 2007, Computing in Science \& Engineering, 9, 90

\bibitem[{{Ivans} {et~al.}(2006){Ivans}, {Simmerer}, {Sneden}, {Lawler},
  {Cowan}, {Gallino}, \& {Bisterzo}}]{Ivans06}
{Ivans}, I.~I., {Simmerer}, J., {Sneden}, C., {et~al.} 2006, \apj, 645, 613

\bibitem[{{Ivans} {et~al.}(2003){Ivans}, {Sneden}, {James}, {Preston},
  {Fulbright}, {H{\"o}flich}, {Carney}, \& {Wheeler}}]{ivans_alphapoor}
{Ivans}, I.~I., {Sneden}, C., {James}, C.~R., {et~al.} 2003, ApJ, 592, 906

\bibitem[{{Jacobson} {et~al.}(2015){Jacobson}, {Keller}, {Frebel}, {Casey},
  {Asplund}, {Bessell}, {Da Costa}, {Lind}, {Marino}, {Norris}, {Pe{\~n}a},
  {Schmidt}, {Tisserand}, {Walsh}, {Yong}, \& {Yu}}]{Jacobson15}
{Jacobson}, H.~R., {Keller}, S., {Frebel}, A., {et~al.} 2015, \apj, 807, 171

\bibitem[{{Ji} \& {Frebel}(2018)}]{ji18}
{Ji}, A.~P., \& {Frebel}, A. 2018, \apj, 856, 138

\bibitem[{{Ji} {et~al.}(2016{\natexlab{a}}){Ji}, {Frebel}, {Chiti}, \&
  {Simon}}]{Ji16a}
{Ji}, A.~P., {Frebel}, A., {Chiti}, A., \& {Simon}, J.~D. 2016{\natexlab{a}},
  \nat, 531, 610

\bibitem[{{Ji} {et~al.}(2016{\natexlab{b}}){Ji}, {Frebel}, {Simon}, \&
  {Chiti}}]{Ji16b}
{Ji}, A.~P., {Frebel}, A., {Simon}, J.~D., \& {Chiti}, A. 2016{\natexlab{b}},
  \apj, 830, 93

\bibitem[{{Johnson}(2002)}]{johnson02a}
{Johnson}, J.~A. 2002, \apjs, 139, 219

\bibitem[{{Johnson} \& {Bolte}(2001)}]{johnson01}
{Johnson}, J.~A., \& {Bolte}, M. 2001, The Astrophysical Journal, 554, 888

\bibitem[{{Johnson} \& {Bolte}(2004)}]{johnson2004}
---. 2004, ApJ, 605, 462

\bibitem[{Jones {et~al.}(2001)Jones, Oliphant, Peterson, {et~al.}}]{scipy}
Jones, E., Oliphant, T., Peterson, P., {et~al.} 2001, {SciPy}: Open source
  scientific tools for {Python}, ,

\bibitem[{{Jonsell} {et~al.}(2006){Jonsell}, {Barklem}, {Gustafsson},
  {Christlieb}, {Hill}, {Beers}, \& {Holmberg}}]{Jonsell06}
{Jonsell}, K., {Barklem}, P.~S., {Gustafsson}, B., {et~al.} 2006, \aap, 451,
  651

\bibitem[{Karakas \& Lattanzio(2014)}]{karakas_lattanzio_2014}
Karakas, A.~I., \& Lattanzio, J.~C. 2014, Publications of the Astronomical
  Society of Australia, 31, e030

\bibitem[{{Keeping}(1962)}]{Keeping62}
{Keeping}, E.~S. 1962, Introduction to Statistical Inference (Princeton, N.J.,
  Van Nostrand)

\bibitem[{{Kelson}(2003)}]{Kelson03}
{Kelson}, D.~D. 2003, \pasp, 115, 688

\bibitem[{{Kilpatrick} {et~al.}(2017){Kilpatrick}, {Foley}, {Kasen},
  {Murguia-Berthier}, {Ramirez-Ruiz}, {Coulter}, {Drout}, {Piro}, {Shappee},
  {Boutsia}, {Contreras}, {Di Mille}, {Madore}, {Morrell}, {Pan}, {Prochaska},
  {Rest}, {Rojas-Bravo}, {Siebert}, {Simon}, \& {Ulloa}}]{kilpatrick17}
{Kilpatrick}, C.~D., {Foley}, R.~J., {Kasen}, D., {et~al.} 2017, Science, 358,
  1583

\bibitem[{{Kirby} {et~al.}(2013){Kirby}, {Cohen}, {Guhathakurta}, {Cheng},
  {Bullock}, \& {Gallazzi}}]{kirby13massmetal}
{Kirby}, E.~N., {Cohen}, J.~G., {Guhathakurta}, P., {et~al.} 2013, \apj, 779,
  102

\bibitem[{{Kratz} {et~al.}(2007){Kratz}, {Farouqi}, {Pfeiffer}, {Truran},
  {Sneden}, \& {Cowan}}]{Kratz07}
{Kratz}, K.-L., {Farouqi}, K., {Pfeiffer}, B., {et~al.} 2007, \apj, 662, 39

\bibitem[{{Kunder} {et~al.}(2017){Kunder}, {Kordopatis}, {Steinmetz},
  {Zwitter}, {McMillan}, {Casagrande}, {Enke}, {Wojno}, \& et~al.}]{kunder17}
{Kunder}, A., {Kordopatis}, G., {Steinmetz}, M., {et~al.} 2017, \aj, 153, 75

\bibitem[{{Kurucz}(1998)}]{kurucz_lines}
{Kurucz}, R.~L. 1998, in IAU Symposium, Vol. 189, Fundamental Stellar
  Properties, ed. T.~R. {Bedding}, A.~J. {Booth}, \& J.~{Davis}, 217

\bibitem[{{Lai} {et~al.}(2008){Lai}, {Bolte}, {Johnson}, {Lucatello}, {Heger},
  \& {Woosley}}]{lai08}
{Lai}, D.~K., {Bolte}, M., {Johnson}, J.~A., {et~al.} 2008, \apj, 681, 1524

\bibitem[{{Li} {et~al.}(2015){Li}, {Zhao}, {Christlieb}, {Wang}, {Wang},
  {Zhang}, {Hou}, \& {Yuan}}]{li15a}
{Li}, H.-N., {Zhao}, G., {Christlieb}, N., {et~al.} 2015, \apj, 798, 110

\bibitem[{{Lind} {et~al.}(2011){Lind}, {Asplund}, {Barklem}, \&
  {Belyaev}}]{Lind11}
{Lind}, K., {Asplund}, M., {Barklem}, P.~S., \& {Belyaev}, A.~K. 2011, \aap,
  528, A103

\bibitem[{{Lindegren} {et~al.}(2018){Lindegren}, {Hern{\'a}ndez}, {Bombrun},
  {Klioner}, {Bastian}, {Ramos-Lerate}, {de Torres}, {Steidelm{\"u}ller},
  {Stephenson}, {Hobbs}, {Lammers}, {Biermann}, {Geyer}, {Hilger}, {Michalik},
  {Stampa}, {McMillan}, {Casta{\~n}eda}, {Clotet}, {Comoretto}, {Davidson},
  {Fabricius}, {Gracia}, {Hambly}, {Hutton}, {Mora}, {Portell}, {van Leeuwen},
  {Abbas}, {Abreu}, {Altmann}, {Andrei}, {Anglada}, {Balaguer-N{\'u}{\~n}ez},
  {Barache}, {Becciani}, {Bertone}, {Bianchi}, {Bouquillon}, {Bourda},
  {Br{\"u}semeister}, {Bucciarelli}, {Busonero}, {Buzzi}, {Cancelliere},
  {Carlucci}, {Charlot}, {Cheek}, {Crosta}, {Crowley}, {de Bruijne}, {de
  Felice}, {Drimmel}, {Esquej}, {Fienga}, {Fraile}, {Gai}, {Garralda},
  {Gonz{\'a}lez-Vidal}, {Guerra}, {Hauser}, {Hofmann}, {Holl}, {Jordan},
  {Lattanzi}, {Lenhardt}, {Liao}, {Licata}, {Lister}, {L{\"o}ffler},
  {Marchant}, {Martin-Fleitas}, {Messineo}, {Mignard}, {Morbidelli}, {Poggio},
  {Riva}, {Rowell}, {Salguero}, {Sarasso}, {Sciacca}, {Siddiqui}, {Smart},
  {Spagna}, {Steele}, {Taris}, {Torra}, {van Elteren}, {van Reeven}, \&
  {Vecchiato}}]{lindegren18}
{Lindegren}, L., {Hern{\'a}ndez}, J., {Bombrun}, A., {et~al.} 2018, \aap, 616,
  A2

\bibitem[{{Marigo} {et~al.}(2017){Marigo}, {Girardi}, {Bressan}, {Rosenfield},
  {Aringer}, {Chen}, {Dussin}, {Nanni}, {Pastorelli}, {Rodrigues}, {Trabucchi},
  {Bladh}, {Dalcanton}, {Groenewegen}, {Montalb{\'a}n}, \& {Wood}}]{Marigo17}
{Marigo}, P., {Girardi}, L., {Bressan}, A., {et~al.} 2017, \apj, 835, 77

\bibitem[{{Mashonkina} {et~al.}(2014){Mashonkina}, {Christlieb}, \&
  {Eriksson}}]{mashonkina14}
{Mashonkina}, L., {Christlieb}, N., \& {Eriksson}, K. 2014, Astronomy and
  Astrophysics, 569, A43

\bibitem[{{Masseron} {et~al.}(2006){Masseron}, {van Eck}, {Famaey}, {Goriely},
  {Plez}, {Siess}, {Beers}, {Primas}, \& {Jorissen}}]{masseron2006}
{Masseron}, T., {van Eck}, S., {Famaey}, B., {et~al.} 2006, \aap, 455, 1059

\bibitem[{{Masseron} {et~al.}(2014){Masseron}, {Plez}, {Van Eck}, {Colin},
  {Daoutidis}, {Godefroid}, {Coheur}, {Bernath}, {Jorissen}, \&
  {Christlieb}}]{Masseron14}
{Masseron}, T., {Plez}, B., {Van Eck}, S., {et~al.} 2014, \aap, 571, A47

\bibitem[{{McWilliam} {et~al.}(1995){McWilliam}, {Preston}, {Sneden}, \&
  {Searle}}]{McWilliametal}
{McWilliam}, A., {Preston}, G.~W., {Sneden}, C., \& {Searle}, L. 1995, AJ, 109,
  2757

\bibitem[{{Mel{\'e}ndez} \& {Barbuy}(2009)}]{mel09}
{Mel{\'e}ndez}, J., \& {Barbuy}, B. 2009, \aap, 497, 611

\bibitem[{{Mel{\'e}ndez} {et~al.}(2016){Mel{\'e}ndez}, {Placco}, {Tucci-Maia},
  {Ram{\'{\i}}rez}, {Li}, \& {Perez}}]{melendez16}
{Mel{\'e}ndez}, J., {Placco}, V.~M., {Tucci-Maia}, M., {et~al.} 2016, \aap,
  585, L5

\bibitem[{{Meyer} {et~al.}(1992){Meyer}, {Mathews}, {Howard}, {Woosley}, \&
  {Hoffman}}]{Meyer92}
{Meyer}, B.~S., {Mathews}, G.~J., {Howard}, W.~M., {Woosley}, S.~E., \&
  {Hoffman}, R.~D. 1992, \apj, 399, 656

\bibitem[{{Nishimura} {et~al.}(2015){Nishimura}, {Takiwaki}, \&
  {Thielemann}}]{Nishimura15}
{Nishimura}, N., {Takiwaki}, T., \& {Thielemann}, F.-K. 2015, \apj, 810, 109

\bibitem[{{O'Brian} {et~al.}(1991){O'Brian}, {Wickliffe}, {Lawler}, {Whaling},
  \& {Brault}}]{obrian91}
{O'Brian}, T.~R., {Wickliffe}, M.~E., {Lawler}, J.~E., {Whaling}, J.~W., \&
  {Brault}, W. 1991, Journal of the Optical Society of America B Optical
  Physics, 8, 1185

\bibitem[{{Placco} {et~al.}(2013){Placco}, {Frebel}, {Beers}, {Karakas},
  {Kennedy}, {Rossi}, {Christlieb}, \& {Stancliffe}}]{placco13}
{Placco}, V.~M., {Frebel}, A., {Beers}, T.~C., {et~al.} 2013, ApJ, 770, 104

\bibitem[{{Placco} {et~al.}(2014){Placco}, {Frebel}, {Beers}, \&
  {Stancliffe}}]{Placco14}
{Placco}, V.~M., {Frebel}, A., {Beers}, T.~C., \& {Stancliffe}, R.~J. 2014,
  \apj, 797, 21

\bibitem[{{Placco} {et~al.}(2015){Placco}, {Beers}, {Ivans}, {Filler}, {Imig},
  {Roederer}, {Abate}, {Hansen}, {Cowan}, {Frebel}, {Lawler}, {Schatz},
  {Sneden}, {Sobeck}, {Aoki}, {Smith}, \& {Bolte}}]{placco15}
{Placco}, V.~M., {Beers}, T.~C., {Ivans}, I.~I., {et~al.} 2015, \apj, 812, 109

\bibitem[{{Placco} {et~al.}(2017){Placco}, {Holmbeck}, {Frebel}, {Beers},
  {Surman}, {Ji}, {Ezzeddine}, {Points}, {Kaleida}, {Hansen}, {Sakari}, \&
  {Casey}}]{placco17}
{Placco}, V.~M., {Holmbeck}, E.~M., {Frebel}, A., {et~al.} 2017, \apj, 844, 18

\bibitem[{{Placco} {et~al.}(2018){Placco}, {Beers}, {Santucci}, {Chanam{\'e}},
  {Sep{\'u}lveda}, {Coronado}, {Points}, {Kaleida}, {Rossi}, {Kordopatis},
  {Lee}, {Matijevi{\v c}}, {Frebel}, {Hansen}, {Holmbeck}, {Rasmussen},
  {Roederer}, {Sakari}, \& {Whitten}}]{Placco18}
{Placco}, V.~M., {Beers}, T.~C., {Santucci}, R.~M., {et~al.} 2018, \aj, 155,
  256

\bibitem[{{Preston} \& {Sneden}(2001)}]{preston_sneden01}
{Preston}, G.~W., \& {Sneden}, C. 2001, AJ, 122, 1545

\bibitem[{{Preston} {et~al.}(2006){Preston}, {Sneden}, {Thompson}, {Shectman},
  \& {Burley}}]{Preston06}
{Preston}, G.~W., {Sneden}, C., {Thompson}, I.~B., {Shectman}, S.~A., \&
  {Burley}, G.~S. 2006, \aj, 132, 85

\bibitem[{{Roederer} {et~al.}(2014{\natexlab{a}}){Roederer}, {Cowan},
  {Preston}, {Shectman}, {Sneden}, \& {Thompson}}]{Roederer14d}
{Roederer}, I.~U., {Cowan}, J.~J., {Preston}, G.~W., {et~al.}
  2014{\natexlab{a}}, \mnras, 445, 2970

\bibitem[{{Roederer} {et~al.}(2018{\natexlab{a}}){Roederer}, {Hattori}, \&
  {Valluri}}]{roederer18d}
{Roederer}, I.~U., {Hattori}, K., \& {Valluri}, M. 2018{\natexlab{a}}, \aj,
  156, 179

\bibitem[{{Roederer} {et~al.}(2016){Roederer}, {Mateo}, {Bailey}, {Spencer},
  {Crane}, \& {Shectman}}]{Roederer16a}
{Roederer}, I.~U., {Mateo}, M., {Bailey}, J.~I., {et~al.} 2016, \mnras, 455,
  2417

\bibitem[{{Roederer} {et~al.}(2014{\natexlab{b}}){Roederer}, {Preston},
  {Thompson}, {Shectman}, {Sneden}, {Burley}, \& {Kelson}}]{Roederer14c}
{Roederer}, I.~U., {Preston}, G.~W., {Thompson}, I.~B., {et~al.}
  2014{\natexlab{b}}, \aj, 147, 136

\bibitem[{{Roederer} {et~al.}(2018{\natexlab{b}}){Roederer}, {Sakari},
  {Placco}, {Beers}, {Ezzeddine}, {Frebel}, \& {Hansen}}]{Roederer2018129}
{Roederer}, I.~U., {Sakari}, C.~M., {Placco}, V.~M., {et~al.}
  2018{\natexlab{b}}, \apj, 865, 129

\bibitem[{{Roederer} {et~al.}(2010){Roederer}, {Sneden}, {Thompson}, {Preston},
  \& {Shectman}}]{roederer10}
{Roederer}, I.~U., {Sneden}, C., {Thompson}, I.~B., {Preston}, G.~W., \&
  {Shectman}, S.~A. 2010, \apj, 711, 573

\bibitem[{{Roederer} {et~al.}(2008){Roederer}, {Frebel}, {Shetrone}, {Allende
  Prieto}, {Rhee}, {Gallino}, {Bisterzo}, {Sneden}, {Beers}, \&
  {Cowan}}]{roederer08}
{Roederer}, I.~U., {Frebel}, A., {Shetrone}, M.~D., {et~al.} 2008, \apj, 679,
  1549

\bibitem[{{Ruffoni} {et~al.}(2014){Ruffoni}, {Den Hartog}, {Lawler}, {Brewer},
  {Lind}, {Nave}, \& {Pickering}}]{ruf14}
{Ruffoni}, M.~P., {Den Hartog}, E.~A., {Lawler}, J.~E., {et~al.} 2014, \mnras,
  441, 3127

\bibitem[{{Sakari} {et~al.}(2018){Sakari}, {Placco}, {Hansen}, {Holmbeck},
  {Beers}, {Frebel}, {Roederer}, {Venn}, {Wallerstein}, {Davis}, {Farrell}, \&
  {Yong}}]{sakari18}
{Sakari}, C.~M., {Placco}, V.~M., {Hansen}, T., {et~al.} 2018, \apjl, 854, L20

\bibitem[{{Sakari} {et~al.}(2019){Sakari}, {Roederer}, {Placco}, {Beers},
  {Ezzeddine}, {Frebel}, {Hansen}, {Sneden}, {Cowan}, {Wallerstein}, {Farrell},
  {Venn}, {Matijevi{\v{c}}}, {Wyse}, {Bland -Hawthorn}, {Chiappini}, {Freeman},
  {Gibson}, {Grebel}, {Helmi}, {Kordopatis}, {Kunder}, {Navarro}, {Reid},
  {Seabroke}, {Steinmetz}, \& {Watson}}]{sakari19}
{Sakari}, C.~M., {Roederer}, I.~U., {Placco}, V.~M., {et~al.} 2019, \apj, 874,
  148

\bibitem[{{Schatz} {et~al.}(2002){Schatz}, {Toenjes}, {Pfeiffer}, {Beers},
  {Cowan}, {Hill}, \& {Kratz}}]{schatz02}
{Schatz}, H., {Toenjes}, R., {Pfeiffer}, B., {et~al.} 2002, ApJ, 579, 626

\bibitem[{{Shappee} {et~al.}(2017){Shappee}, {Simon}, {Drout}, {Piro},
  {Morrell}, {Prieto}, {Kasen}, {Holoien}, {Kollmeier}, {Kelson}, {Coulter},
  {Foley}, {Kilpatrick}, {Siebert}, {Madore}, {Murguia-Berthier}, {Pan},
  {Prochaska}, {Ramirez-Ruiz}, {Rest}, {Adams}, {Alatalo}, {Ba{\~n}ados},
  {Baughman}, {Bernstein}, {Bitsakis}, {Boutsia}, {Bravo}, {Di Mille}, {Higgs},
  {Ji}, {Maravelias}, {Marshall}, {Placco}, {Prieto}, \& {Wan}}]{shappee17}
{Shappee}, B.~J., {Simon}, J.~D., {Drout}, M.~R., {et~al.} 2017, Science, 358,
  1574

\bibitem[{Siegel {et~al.}(2019)Siegel, Barnes, \&
  Metzger}]{siegel_collapsars_2019}
Siegel, D.~M., Barnes, J., \& Metzger, B.~D. 2019, Nature, 569, 241

\bibitem[{{Siqueira Mello} {et~al.}(2012){Siqueira Mello}, {Barbuy}, {Spite},
  \& {Spite}}]{mello12}
{Siqueira Mello}, C., {Barbuy}, B., {Spite}, M., \& {Spite}, F. 2012, \aap,
  548, A42

\bibitem[{{Siqueira Mello} {et~al.}(2014){Siqueira Mello}, {Hill}, {Barbuy},
  {Spite}, {Spite}, {Beers}, {Caffau}, {Bonifacio}, {Cayrel}, {Fran{\c c}ois},
  {Schatz}, \& {Wanajo}}]{mello}
{Siqueira Mello}, C., {Hill}, V., {Barbuy}, B., {et~al.} 2014, \aap, 565, A93

\bibitem[{{Sk{\'u}lad{\'o}ttir} {et~al.}(2015){Sk{\'u}lad{\'o}ttir}, {Tolstoy},
  {Salvadori}, {Hill}, {Pettini}, {Shetrone}, \& {Starkenburg}}]{sku15}
{Sk{\'u}lad{\'o}ttir}, {\'A}., {Tolstoy}, E., {Salvadori}, S., {et~al.} 2015,
  \aap, 574, A129

\bibitem[{{Sneden} {et~al.}(2008){Sneden}, {Cowan}, \& {Gallino}}]{Sneden08}
{Sneden}, C., {Cowan}, J.~J., \& {Gallino}, R. 2008, \araa, 46, 241

\bibitem[{{Sneden} {et~al.}(2016){Sneden}, {Cowan}, {Kobayashi}, {Pignatari},
  {Lawler}, {Den Hartog}, \& {Wood}}]{Sneden16}
{Sneden}, C., {Cowan}, J.~J., {Kobayashi}, C., {et~al.} 2016, \apj, 817, 53

\bibitem[{{Sneden} {et~al.}(2009){Sneden}, {Lawler}, {Cowan}, {Ivans}, \& {Den
  Hartog}}]{Sneden09}
{Sneden}, C., {Lawler}, J.~E., {Cowan}, J.~J., {Ivans}, I.~I., \& {Den Hartog},
  E.~A. 2009, \apjs, 182, 80

\bibitem[{{Sneden} {et~al.}(2014){Sneden}, {Lucatello}, {Ram}, {Brooke}, \&
  {Bernath}}]{sneden14}
{Sneden}, C., {Lucatello}, S., {Ram}, R.~S., {Brooke}, J.~S.~A., \& {Bernath},
  P. 2014, \apjs, 214, 26

\bibitem[{Sneden {et~al.}(1996)Sneden, McWilliam, Preston, Cowan, Burris, \&
  Amorsky}]{Snedenetal:1996}
Sneden, C., McWilliam, A., Preston, G.~W., {et~al.} 1996, ApJ, 467, 819

\bibitem[{{Sneden}(1973)}]{Sneden73}
{Sneden}, C.~A. 1973, PhD thesis, The University of Texas at Austin.

\bibitem[{{Sobeck} {et~al.}(2011){Sobeck}, {Kraft}, {Sneden}, {Preston},
  {Cowan}, {Smith}, {Thompson}, {Shectman}, \& {Burley}}]{Sobeck11}
{Sobeck}, J.~S., {Kraft}, R.~P., {Sneden}, C., {et~al.} 2011, \aj, 141, 175

\bibitem[{{Spite} {et~al.}(2014){Spite}, {Spite}, {Bonifacio}, {Caffau},
  {Fran{\c c}ois}, \& {Sbordone}}]{spite14}
{Spite}, M., {Spite}, F., {Bonifacio}, P., {et~al.} 2014, \aap, 571, A40

\bibitem[{{Surman} {et~al.}(2006){Surman}, {McLaughlin}, \& {Hix}}]{surman06}
{Surman}, R., {McLaughlin}, G.~C., \& {Hix}, W.~R. 2006, The Astrophysical
  Journal, 643, 1057

\bibitem[{{Tody}(1986)}]{irafa}
{Tody}, D. 1986, in \procspie, Vol. 627, Instrumentation in astronomy VI, ed.
  D.~L. {Crawford}, 733

\bibitem[{{Tody}(1993)}]{irafb}
{Tody}, D. 1993, in Astronomical Society of the Pacific Conference Series,
  Vol.~52, Astronomical Data Analysis Software and Systems II, ed. R.~J.
  {Hanisch}, R.~J.~V. {Brissenden}, \& J.~{Barnes}, 173

\bibitem[{{Travaglio} {et~al.}(2004){Travaglio}, {Gallino}, {Arnone}, {Cowan},
  {Jordan}, \& {Sneden}}]{Travaglio04}
{Travaglio}, C., {Gallino}, R., {Arnone}, E., {et~al.} 2004, \apj, 601, 864

\bibitem[{{Truran} {et~al.}(2002){Truran}, {Cowan}, {Pilachowski}, \&
  {Sneden}}]{truran02}
{Truran}, J.~W., {Cowan}, J.~J., {Pilachowski}, C.~A., \& {Sneden}, C. 2002,
  PASP, 114, 1293

\bibitem[{{van~der~Walt} {et~al.}(2011){van~der~Walt}, Colbert, \&
  Varoquaux}]{numpy}
{van~der~Walt}, S., Colbert, S.~C., \& Varoquaux, G. 2011, Computing in Science
  \& Engineering, 13, 22

\bibitem[{{Vasiliev}(2019)}]{vasiliev19}
{Vasiliev}, E. 2019, \mnras, 482, 1525

\bibitem[{{Walker} {et~al.}(2016){Walker}, {Mateo}, {Olszewski}, {Koposov},
  {Belokurov}, {Jethwa}, {Nidever}, {Bonnivard}, {Bailey}, {Bell}, \&
  {Loebman}}]{walker16}
{Walker}, M.~G., {Mateo}, M., {Olszewski}, E.~W., {et~al.} 2016, \apj, 819, 53

\bibitem[{{Wanajo}(2013)}]{Wanajo13}
{Wanajo}, S. 2013, \apjl, 770, L22

\bibitem[{{Wanajo} \& {Ishimaru}(2006)}]{wanajo_ishimaru}
{Wanajo}, S., \& {Ishimaru}, Y. 2006, Nuclear Physics A, 777, 676

\bibitem[{{Wanajo} {et~al.}(2001){Wanajo}, {Kajino}, {Mathews}, \&
  {Otsuki}}]{wanajo01}
{Wanajo}, S., {Kajino}, T., {Mathews}, G.~J., \& {Otsuki}, K. 2001, ApJ, 554,
  578

\bibitem[{{Woosley} \& {Hoffman}(1992)}]{Woosley92}
{Woosley}, S.~E., \& {Hoffman}, R.~D. 1992, \apj, 395, 202

\bibitem[{{Yong} {et~al.}(2013){Yong}, {Norris}, {Bessell}, {Christlieb},
  {Asplund}, {Beers}, {Barklem}, {Frebel}, \& {Ryan}}]{yong13_II}
{Yong}, D., {Norris}, J.~E., {Bessell}, M.~S., {et~al.} 2013, ApJ, 762, 26

\bibitem[{{Yuan}} {et~al.}(2020){Yuan}, {Myeong}, {Beers}, {Evans}, {Lee}, {Banerjee}, {Gudin}, {Hattori}, {Li}, {Matsuno}]{yuan20} {Yuan}, Z., {Myeong}, G.~C., {Beers}, T.~C., {et~al.} 2020, ApJ, 891, 1

\end{thebibliography}
\end{document}